\documentclass{bmcart}

\RequirePackage{hyperref}
\usepackage[utf8]{inputenc} 

\usepackage{hyperref}
\usepackage{booktabs}
\usepackage[export]{adjustbox}
\usepackage{graphicx}
\usepackage{graphbox}
\usepackage{subfigure}
\usepackage{longtable}

\begin{document}
\begin{frontmatter}

\begin{fmbox}
\dochead{Research}

\title{Safe Spaces or Toxic Places?  Content Moderation and Social Dynamics of Online Eating Disorder Communities
}
\author[
   addressref={aff1,aff2},
   corref={aff1},                       
   email={lerman@isi.edu}
]{\inits{KL}\fnm{Kristina} \snm{Lerman}}
\author[
   addressref={aff1, aff2},              
   email={mhchu@usc.edu}   
]{\inits{MDC}\fnm{Minh Duc} \snm{Chu}}
\author[
   addressref={aff1,aff2},
   email={}
]{\inits{CB}\fnm{Charles} \snm{Bickham}}
\author[
   addressref={aff1,aff2},
   email={}
]{\inits{LL}\fnm{Luca} \snm{Luceri}}
\author[
   addressref={aff1,aff2},
   email={}
]{\inits{EF}\fnm{Emilio} \snm{Ferrara}}

\address[id=aff1]{
  \orgname{Information Sciences Institute, University of Southern California}, 
  \city{Marina del Rey, CA},                              
  \cny{USA}                                    
}

\address[id=aff2]{%
  \orgname{Thomas Lord Department of Computer Science, University of Southern California},
  \city{Los Angeles, CA},
  \cny{USA}
}

\end{fmbox}

\begin{abstractbox}
\begin{abstract}
Social media platforms have become critical spaces for discussing mental health concerns, including eating disorders. While these platforms can provide valuable support networks, they may also amplify harmful content that glorifies disordered cognition and self-destructive behaviors.
While social media platforms have implemented various content moderation strategies, from stringent to laissez-faire approaches, we lack a comprehensive understanding of how these different moderation practices interact with user engagement in online communities around these sensitive mental health topics.
This study addresses this knowledge gap through a comparative analysis of eating disorder discussions across Twitter/X, Reddit, and TikTok. Our findings reveal that while users across all platforms engage similarly in expressing concerns and seeking support, platforms with weaker moderation (like Twitter/X) enable the formation of toxic echo chambers that amplify pro-anorexia rhetoric. These results demonstrate how moderation strategies significantly influence the development and impact of online communities, particularly in contexts involving mental health and self-harm.


\end{abstract}

\end{abstractbox}

\end{frontmatter}




\section{Introduction}
\textcolor{red}{[\textbf{Warning: This paper discusses eating disorders. Reader discretion is advised.}]}

Online communities provide a vital space for socializing, sharing information, and receiving social and emotional  support~\cite{ellison2007benefits}. However, maintaining safe, welcoming, and inclusive online environments has proven challenging. Online communities that indoctrinate users into extreme ideologies~\cite{ferrara2016predicting, ferrara2017contagion,  schmitz2022quantifying, wang2022identifying, zimmerman2024attraction} or glorify self-harm~\cite{Goldenberg2022ncri} are especially challenging to moderate, as users often evade moderation through the use of coded language, insider jargon, and intentional misspellings~\cite{chancellor2016thyghgapp, cobb2017not, bickham2024hidden}.
While some social media platforms have made progress in identifying and moderating online speech that violates community norms, such as hate speech and personal attacks~\cite{jhaver2019human, rajadesingan2020quick}, others have adopted a more laissez-faire approach.
These varied strategies for moderating online content, including discussions on sensitive topics like body image and mental health, have produced vastly different user experiences across platforms.

This paper examines online discussions of eating disorders on different social media platforms as a microcosm of mental health concerns.  
Eating disorders represent a serious mental health condition characterized by obsessive thoughts and unhealthy behaviors around food, eating, and body size. The condition, which includes anorexia, bulimia, and binge eating disorder, affects 24 million people in the US ~\cite{vanHoeken2020review}
and is among the deadliest of all mental health conditions,  with over 10,000 deaths each year in the US. 
The COVID-19 pandemic has amplified the disorder~\cite{nutley2021impact}, particularly among girls and young women~\cite{Hartman2022}. According to the Centers for Disease Control and Prevention (CDC), the number of weekly visits to emergency rooms for adolescent girls struggling with eating disorders has doubled in 2021 compared to 2019~\cite{radhakrishnan2022pediatric}.

Researchers have historically linked the rise in eating disorders to the proliferation of the ``thin ideal'' images in the media~\cite{becker2004television} and more recently on social media platforms like Instagram and TikTok~\cite{fardouly2016social, marks2020pursuit, lavender2017men}. Exposure to idealized body images invites negative comparisons and pressure to conform to unrealistic beauty standards, which contribute to negative body image, a known risk factor for developing eating disorders~\cite{choukas2022perfectstorm}.

However, the relationship between social media and eating disorders extends beyond mere exposure to such imagery and is connected to the social dynamics of online platforms. Social media algorithms direct vulnerable individuals searching for weight loss or fitness advice toward harmful content promoting extreme dieting and over-exercising. Additionally, users may encounter pro-anorexia communities or influencers who glorify thinness and present anorexia as an aesthetic ideal rather than a serious medical condition~\cite{Ging2018, wang2018social}.
While interactions within online communities can provide emotional support and validation~\cite{oksanen2016proanorexia, YeshuaKatz2013stigma}, unsafe communities may ultimately harm their members by normalizing disordered thinking and delaying recovery~\cite{pater2016hunger, pater2017defining, chancellor2016recovery}. Without robust content moderation policies, these platforms risk exacerbating the problem by enabling the creation and perpetuation of harmful content, reinforcing toxic social dynamics, and deepening the impact on vulnerable users. Effective moderation is critical to disrupting these dynamics and fostering healthier online environments.


To explore the interplay between community moderation and social dynamics, we conduct a comparative analysis of eating disorder discussions on Twitter, Reddit, and TikTok. 
These platforms vary in their approaches to moderating eating disorder content. TikTok redirects searches for pro-anorexia content to mental health resources.  Twitter has historically taken a more lenient stance, allowing content glorifying eating disorders to proliferate~\cite{Goldenberg2022ncri}, while Reddit, which relies on community moderators,  banned pro-anorexia forums since 2018.

We collected posts from TikTok and Twitter using the same set of search terms related to dieting, weight loss, body image concerns, and eating disorders, and we gathered relevant discussions from Reddit forums on such themes. Our analysis reveals that users on all platforms discuss body image and eating disorder concerns, often integrating these topics into broader conversations about dieting and fitness. We also measure emotions and toxicity expressed in discussions. Our results show that across platforms, users engage with online communities in a similar manner, venting their negative feelings and receiving emotional support through positive interactions.

However, among the three platforms, Twitter stands out for its toxicity. Semantically, content that glorifies anorexia and self-harm proliferates on the site. By contrast, eating disorder content on the other platforms is more recovery-oriented, while also being more integrated within mainstream topics. Socially, users link within tightly-knit communities on Twitter that are isolated from others. In contrast, eating disorders communities on Reddit are integrated with other topics.

Our findings show that weaker moderation on platforms like Twitter can enable harmful communities to flourish, creating echo chambers that amplify pro-anorexia rhetoric and ensnare ever more vulnerable users. This vicious spiral, similar to online  radicalization~\cite{Kruglanski2022}, provides a plausible mechanism for the increase in content glorifying eating disorders on Twitter~\cite{Goldenberg2022ncri}. 

Our work suggests that effective guardrails prevent the generation of harmful content without compromising the ability of online communities to provide emotional support. 
Understanding the interplay between online communities and moderation practices is crucial to minimizing social media’s negative impact on mental health and promoting recovery-oriented discussions.

\section{Related Works}



Eating disorders have complex biopsychosocial etiology, with contributing biological factors~\cite{aman2022prevalence} and psychological comorbidities, such as anxiety and perfectionism.
In addition, social mechanisms like peer effects~\cite{allison2014anorexia} and exposure to idealized body images in the media contribute to eating disorders. For instance, before the introduction of Western TV programming in Fiji in 1995, purging to control weight was virtually unknown, but the practice grew quickly  afterwards~\cite{becker2004television}.

\paragraph{Social Media and Body Image Concerns.}
Social media fuels body image concerns, a key risk factor for developing depression and eating disorders~\cite{choukas2022perfectstorm}. Content on image-based platforms like Instagram, Snapchat, and TikTok features images of often heavily edited bodies that promote the ``thin ideal'' or the ``muscular ideal''. 
Adolescents are often not aware that images are heavily edited~\cite{marks2020pursuit}, and as a result, they normalize distorted, unrealistic beauty ideals. 
Exposure to the ``thin ideal'' or the ``muscular ideal'' can provoke upward social comparisons, where individuals compare themselves to people whom they believe to be better than themselves~\cite{harriger2022rabbithole} and as a result feel worse about their own appearance~\cite{Saiphoo2019, fardouly2016social,choukas2022perfectstorm}.
Social media can further fuel body image concerns through negative feedback to users' content and the images posted by others. In extreme cases, the negative feedback can manifest as body shaming and cyberbullying.

\paragraph{Online Pro-anorexia Communities.}
Pro-ana communities are online spaces, like blogs and online forums,  that promote anorexia as a lifestyle, and not as an illness. Pro-ana communities provide a venue for individuals with eating disorders to share tips on losing weight and concealing weight loss from others, as well as ``thinspiration'' images of very thin bodies to motivate weight loss~\cite{Ging2018,oksanen2015pro}.
Researchers have argued that pro-ana communities have both positive and negative effects on individuals with eating disorders.
Such communities provide social support~\cite{juarascio2010pro} and a sense of belonging to individuals who often feel stigmatized and misunderstood~\cite{oksanen2016proanorexia,YeshuaKatz2013stigma}. Members can find empathy, encouragement, a safe space to vent, and information to help them better understand and manage their illness~\cite{McCormack2010}. 
On the negative side, pro-ana communities often promote unhealthy behaviors, such as extreme dietary restrictions and over-exercising. Members may compete with each other in weight loss, ask the group to hold them accountable to their weight loss goals, or find ``buddies'' to go through the same difficult periods of food restriction. Combined with celebrating the ``thin ideal'', these communities can increase psychological distress around body image and exacerbate eating disorders~\cite{Mento2021}. 
The dual nature of such communities, offering benefits and risks to vulnerable individuals struggling with eating disorders, highlights the challenge of moderating online communities.

Computer scientists have examined the language used by pro-anorexia communities to identify problematic content or users at risk. 
\cite{chancellor2016post} build a lexical classifier to identify posts that will be taken down for violating Instagram's rules against self-harm. 
The same authors compared pro-recovery and pro-anorexia communities to show how people move from illness to recovery~\cite{chancellor2016recovery}. In contrast, rather than focus on individuals, we aim to identify the social mechanisms, which can be targeted to disrupt the growth of pro-anorexia communities.

\cite{wang2018social} analyzed pro-anorexia communities from a network perspective, examining the retweet network of people discussing eating disorders and their emotions and using hashtags to classify attitudes toward eating disorders or recovery. 

\paragraph{Community Moderation on Social Media.}

Community moderation aims to create safe and supportive spaces in online communities while curbing toxicity, harmful and unsafe content. This is particularly challenging in sensitive health discussions, where overly lenient or strict moderation can have lasting consequences. Research shows that moderated mental health discussions can lead to greater user participation, increased civility, and improved emotional outcomes~\cite{wadden2021effect}. 
However, extremely strict moderation in sensitive health communities---like those discussing eating disorders---can inadvertently silence personal narratives and remove vital support~\cite{feuston2020conformity}.

Social media platforms employ a range of content moderation strategies to manage mental health communities. Traditional methods often involve censoring or banning specific unsafe keywords and hashtags, such as ``depression'', ``suicide'', or ``proana''~\cite{zhang2024debate}. Platforms like Instagram and Tumblr implement these bans and also direct users to mental health resources when they search for or post related content. Reddit, in contrast, relies on a decentralized model where community moderators specify and enforce community rules. When this system fails, Reddit may intervene by banning or quarantining harmful communities, as it did in 2018 when it banned subreddits glorifying anorexia. 

Despite advances in automated detection of harmful content, including toxicity and hate speech~\cite{davidson2017automated}, these tools often lack the nuance needed for effective moderation and risk over-blocking~\cite{chandrasekharan2019crossmod}, disproportionately affecting vulnerable or marginalized groups~\cite{dorn2023non}. Additionally, rapidly evolving community jargon can render such tools ineffective~\cite{chancellor2016thyghgapp,bickham2024hidden}.

Our work differs from existing literature in that it explores the social dynamics of online eating disorders communities on various social media platforms, focusing on commonalities, like community formation processes and emotional engagement. 
Despite differences in platform design and content moderation policies, our work identifies similarities in eating disorder discussions and behaviors across different social media platforms.

\section{Data and Methods}

\begin{figure}
    \centering
    \begin{tabular}{ccc}
    \includegraphics[width=0.25\linewidth]{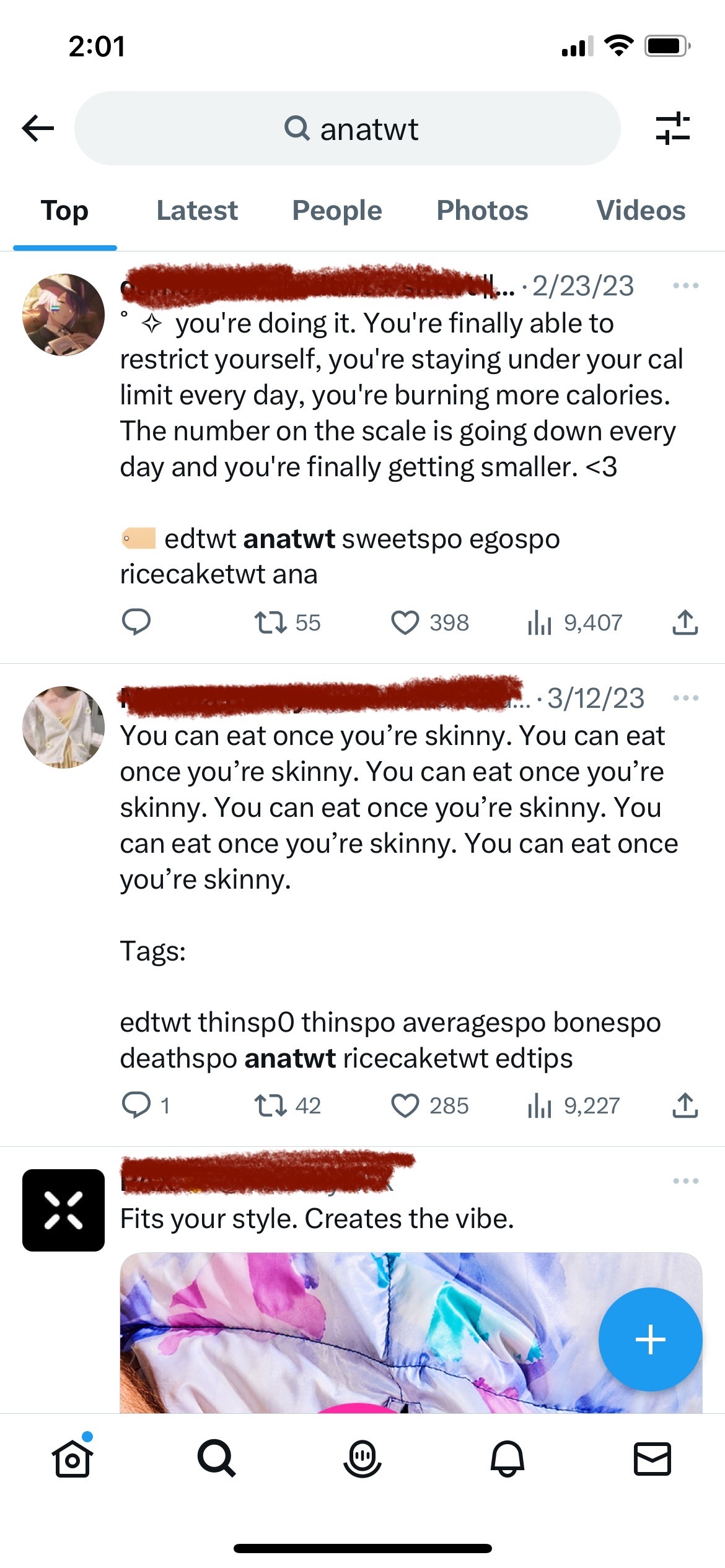} &
    \includegraphics[width=0.25\linewidth]{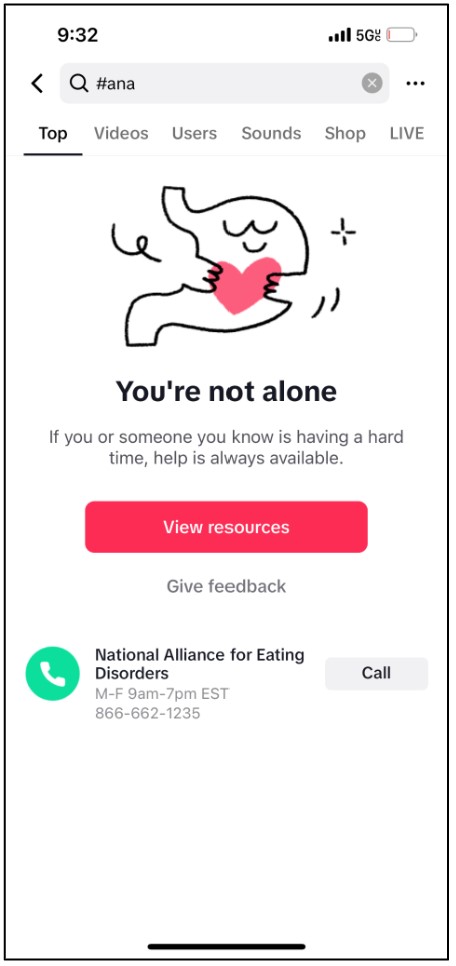} &
    \includegraphics[width=0.25\linewidth]{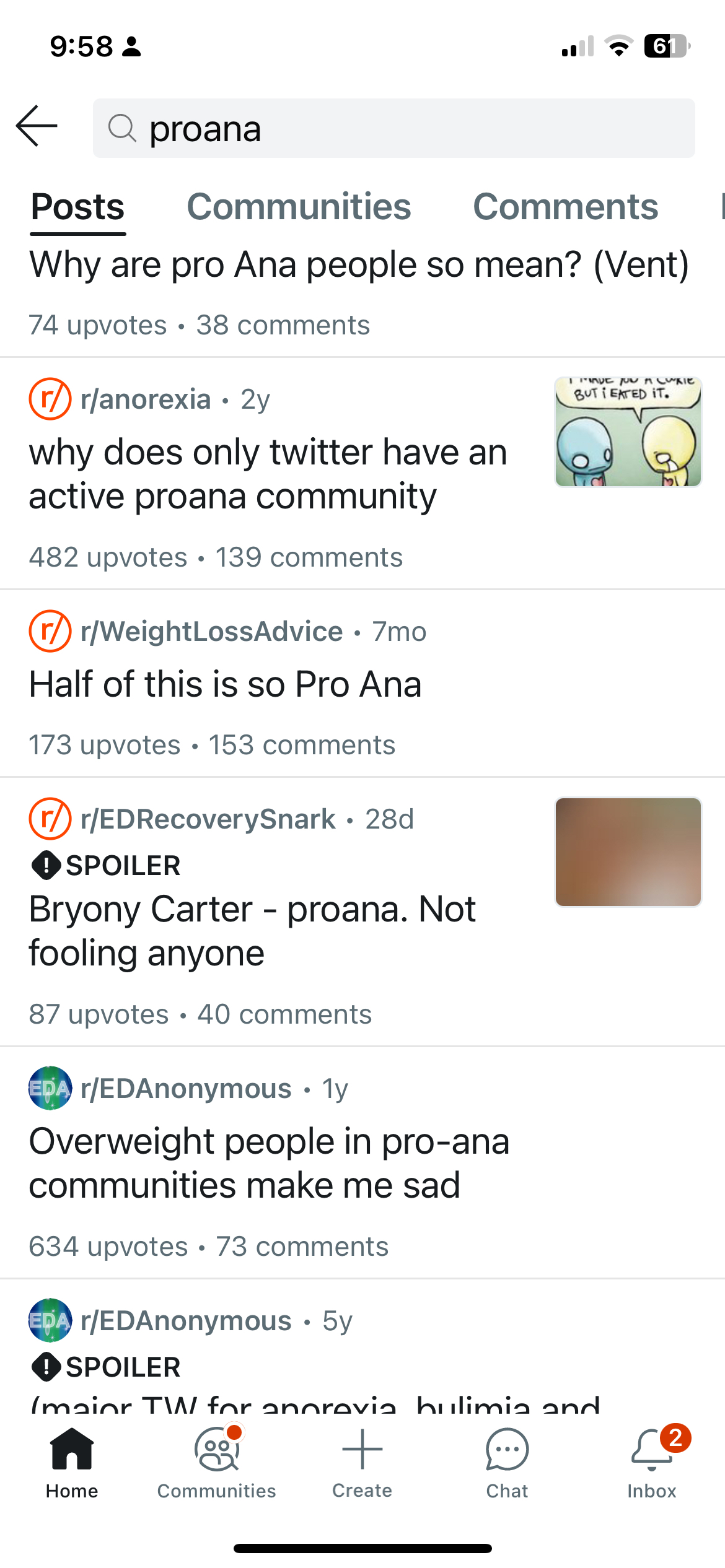} \\
    (a) Twitter & (b) TikTok  & (c) Reddit      \end{tabular}
    \caption{Screencaps of searches for anorexia-related content. (a) Searching for ``anatwt'' a term used by the pro-anorexia community on Twitter, returned posts promoting food restriction. (b) Search for ``ana'' was blocked on TikTok, redirecting users to mental health resources. (c) Search for ``proana'' on Reddit directs users to discussions in different forums.}
    \label{fig:screencaps}
\end{figure}

\subsection{Data}
\label{sec:data}

The platforms we study provide spaces for users to share personal updates and engage with others, but each platform’s unique features shape the nature of these interactions.

\paragraph{Twitter.}
On Twitter (now known as X), users follow accounts to see their updates, creating communities of ``mutual followers'' or ``moots.'' Despite policies prohibiting content that glorifies suicide and self-harm, such content has proliferated. A study found a 500\% increase in posts mentioning self-harm between October 2021 and August 2022~\cite{Goldenberg2022ncri}, averaging 20,000 posts per month. Content moderation on Twitter relies heavily on user reports and automated systems to flag violations. Although some eating disorder communities self-regulate using content trigger warnings (e.g., ``TW: eating disorder''), Twitter's broader moderation remains lax, with search terms like ``anatwt,'' associated with pro-anorexia content, still unflagged as of March 2023 (Fig.~\ref{fig:screencaps}(a)).

We collected 2.6M tweets from 557K users using keywords to query Twitter for messages covering the period from 2022-10 to 2023-03\footnote{The collection was done in March 2023.}. 
Our starting point was a set of terms identified by prior works~\cite{chancellor2016post,pater2016hunger} as promoting eating disorders. This include \textit{thinspo} (``thininspiration''), \textit{proana} (pro-anorexia), and \textit{pro-mia} (pro-bulimia), among others. After examining results, we removed terms like \textit{skinny}, which returned links to pornographic sites or product advertisements and added terms related to diet and weight loss topics such as (\textit{ketodiet}, \textit{weightloss}, $\ldots$), and anti-diet culture  (\textit{bodypositivity}, \textit{dietculture}, $\ldots$). See the Appendix for the full list of keywords. Information collected on posts includes user profile descriptions, timestamps, message content, and hashtags.

\paragraph{TikTok.}
TikTok emphasizes personal video content, with algorithms curating user timelines. Interactions occur through comments on videos, and moderation relies on user reports alongside automated systems that flag potential violations for human review \cite{TikTok2024moderation, lookingbill2024there}. TikTok has ramped up efforts to block harmful content related to eating disorders, with reported violations rising from 0.087\% in 2023 to 3.669\% in early 2024 \cite{TikTok2024governmentremoval}. Hashtags like \#ana are no longer searchable, and users are redirected to mental health resources (Fig.~\ref{fig:screencaps}(b)). As of May 2024, TikTok updated its guidelines to ban the promotion of disordered eating and dangerous weight loss behaviors, restricting such content for users under 18 in the ``For You Feed'' \cite{TikTok2024disorderedeating}.

We collected 14,816 video posts from 6,612 TikTok users using the same keywords as Twitter.  The dataset spans from December 2016 to April 2023 and includes over 562,856 comments.
Each video includes textual information, captured through the video description (Description) and a list of hashtags used in the video (Challenges). Additionally, the dataset includes the comments the videos received.

\paragraph{Reddit.}
Reddit’s moderation combines automated systems and human moderators to enforce platform-wide and subreddit-specific rules. In 2018, Reddit banned pro-eating disorder subreddits like \texttt{r/ProED}. In addition to banning, Reddit uses ``quarantining'' to isolate sensitive communities with warning screens and remove them from public feeds. Reddit’s decentralized model allows human moderators to manage subreddits using tools like Automoderator~\cite{jhaver2019human} and machine learning-based methods~\cite{he2023cpl}. The platform’s structure, with nested comments and threaded discussions, supports in-depth, targeted conversations.

We collected Reddit data via Academic Torrent.\footnote{https://academictorrents.com/}, which compiles submissions and comments using the Pushshift API \cite{Baumgartner_Zannettou_Keegan_Squire_Blackburn_2020} Our data, covering January 2019 to November 2023, includes 46 subreddits related to fitness, diet, eating disorders, and mental health.

We identified these relevant subreddits by combining keyword searches based on our expertise and literature with Reddit's search recommendations and found 26 subreddits. After verifying the activity and relevance of those subreddits, we build a network of subreddit mentions from submissions and comments. This revealed additional subreddits frequently mentioned by the initial set. Using the Louvain algorithm, we identified community clusters and added 28 highly connected subreddits, bringing the total to 54.

We filtered bot-generated content (e.g., \textit{AutoModerator, steroidsBot, EDAnonymous\_Bot}), removed duplicates, and discarded subreddits with fewer than 500 submissions, reducing the list to 46 subreddits (Table \ref{tab:subreddits}). To balance data across subreddits, we randomly sampled up to 5,000 submissions and 5,000 comments per subreddit, resulting in 178,272 submissions, 218,139 comments, and 212,529 unique users.

\subsection{Network Analysis}
\label{sec:method_net_anlysis}
Research on social media suggests that users tend to exchange information and engage with others who share similar traits or interests \cite{stewart2018examining, rmit}. By constructing interaction-based networks from our data, we can uncover organically formed online communities focused on discussing eating disorders.

\paragraph{Retweet Network on Twitter.}
On Twitter (or X), retweeting plays a central role in spreading information, functioning as a means of content distribution \cite{suh2010want} and a proxy of attention via following~\cite{rao2023retweets}. To capture this dynamic, we construct a retweet network, where each node represents a user, and edges signify interactions between users who have retweeted each other at least once.

TikTok users engage by liking or commenting on videos posted by others, rather than sharing them with their own followers. While previous studies constructed social networks based on these likes \cite{bonifazi2022investigating} or replies \cite{basch2021community},  we are unable to do so since we did not collect individual interaction data. In future work, we will expand data collection to study TikTok communities.

\paragraph{Mention Network on Reddit.}
In contrast to Twitter, Reddit lacks typical social networking features like friendships or followers. Instead, it is organized around subreddits—smaller communities of users with shared interests where they interact with each other frequently. Given the high level of homophily within subreddits, we treat each subreddit as a distinct unit. To examine the network structure on Reddit, we construct a directed network of subreddit mentions, where each node represents a subreddit, and an edge indicates an interaction where a user mentions another subreddit in their post title, content, or comments. The weight of each edge reflects the frequency of these interactions. Using regular expressions \cite{aho1991algorithms}, we extract subreddit mentions by identifying strings that match the pattern \texttt{r/{subreddit\_name}}. Each mention within a post or comment is counted once. To be included in the analysis, subreddits must be mentioned at least ten times, including open, banned, and quarantined subreddits. 



\paragraph{Hashtag Co-occurrence Network.}
We create a \textit{hashtag co-occurrence network}, which represents relationships between topics that users choose to highlight with a ``\#'' symbol. An edge between two hashtags exists if they appear together in a post, with the weight of the edge representing the number of times--i.e., the number of posts in which---the two hashtags co-occur. 

\paragraph{Community Detection.}
We use community detection (specifically Louvain modularity maximization~\cite{Blondel_2008}) to identify tightly-knit clusters of nodes that characterize the structure of the network. The clusters in the retweet networks represent communities of like-minded users who frequently retweet each other. 
Similarly, dense clusters within the hashtag co-occurrence network are semantically related topics that are frequently mentioned together.


\subsection{Measuring Emotions and Toxicity}

Emotions play a vital role in social interactions and are crucial for fostering supportive and safe online environments~\cite{prescott2019young}. Language in online communication carries cues to discrete emotions, including positive emotions, like joy and love, and negative emotions like anger and disgust. 

To detect emotions in tweets and TikTok text, we utilize SpanEmo~\cite{alhuzali2021spanemo}, a state-of-the-art emotion detection model trained on the SemEval 2018 Task 1 E-c dataset~\cite{SemEval2018Task1}. This dataset includes posts labeled with anger, anticipation, disgust, fear, joy, optimism, pessimism, sadness, surprise, trust, and neutral (no emotion). SpanEmo, which uses BERT \cite{devlin2018bert} to encode text, outputs a continuous confidence score for each emotion, indicating its likelihood of being present. The model can predict multiple emotions or none for a given tweet, classifying text into 11 emotions as described. 

For analyzing emotions in Reddit posts, we use the model trained on the GoEmotions dataset \cite{demszky2020goemotions}, a multilabel classification model available on HuggingFace.\footnote{https://huggingface.co/SamLowe/roberta-base-go\_emotions} This dataset consists of Reddit comments labeled with 28 categories, including 27 emotions and a neutral category. The model produces a score between 0 and 1, representing the confidence for each emotion. To align with SpanEmo’s categories, we map the closest GoEmotions equivalents: “Excitement” for “Anticipation,” “Disappointment” for “Pessimism,” and “Admiration” for “Trust.”

For detecting toxicity, we employ Detoxify \cite{Detoxify}, a model designed to quantify toxic language in online interactions. It provides scores across six categories of toxicity, including general toxicity (used in this study), severe toxicity, obscenity, threats, insults, identity attacks, and sexually explicit content.

\section{Results}
We use the methods described above to analyze eating disorders-related content on social media platforms. We examine the structure of the topic space and interactions between users and go beyond content to examine emotional expressions and engagement.

\subsection{Thematic Organization of Eating Disorder Content}
\label{sec:h1}

\begin{figure*}[htp]
    \centering
\begin{tabular}{c}
    \includegraphics[width=0.7\linewidth]{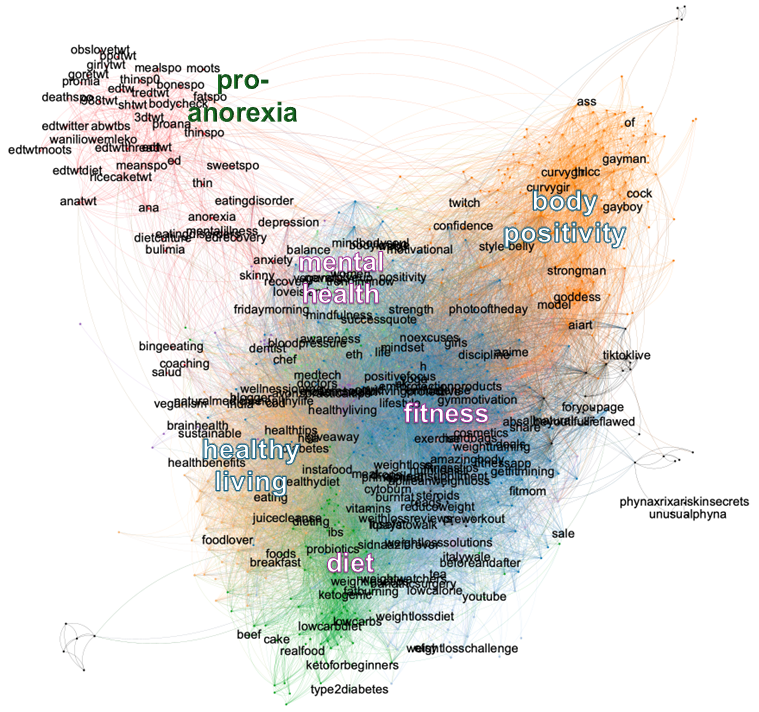} \\
    (a) Twitter \\
    \includegraphics[width=0.7\linewidth]{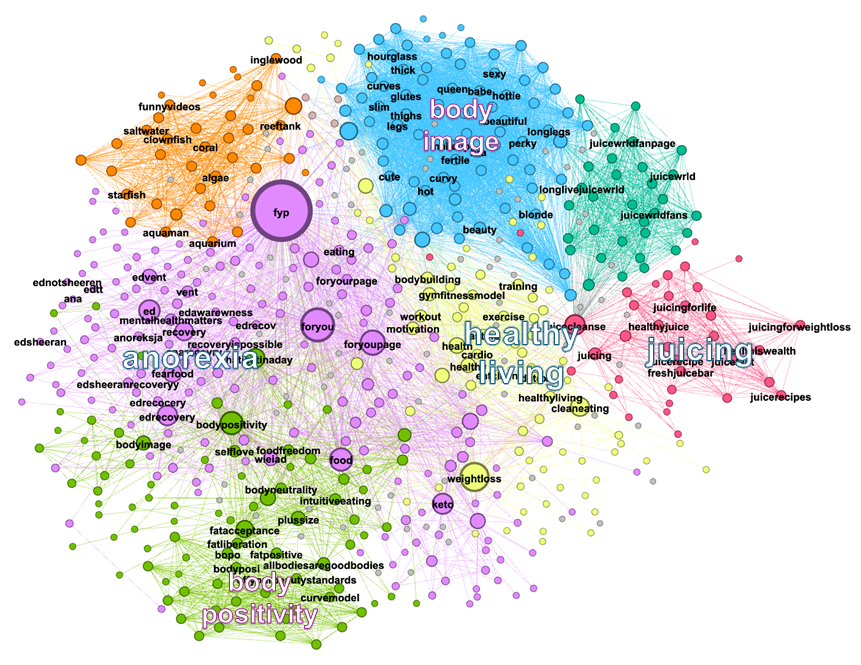} \\
    (b) TikTok
\end{tabular}
    \caption{Hashtag co-occurrence network on (top) Twitter and (bottom) TikTok. Nodes are popular hashtags and edges link hashtags that are frequently used together. Node colors represent discovered communities and 15\% of the labels are shown.}
    \label{fig:hashtag-network}
\end{figure*}

Social media platforms offer different means for users to organize content. On Twitter and TikTok, for example, users use hashtags as keywords to highlight important topics. By looking the relationships between hashtags, we can analyze the global thematic structure to understand how the content related to eating disorders is integrated into mainstream dieting and weight loss discussions.

Figure~\ref{fig:hashtag-network} shows the topic landscape on Twitter and TikTok in the form of the hashtag co-occurrence network of popular hashtags, with edges linking hashtags that frequently appear together in the same post. 
Community detection identifies dense clusters of hashtags (denoted by different colors in Fig.~\ref{fig:hashtag-network}) that represent related topics like diet and weight loss, fitness and healthy living, nutrition and recipes, body positivity and self-acceptance, etc. 

There are many similarities in the global organization of content on Twitter and TikTok. Both platforms have similar topic clusters, like ``healthy living'', which includes Twitter hashtags \#healthylife, \#vegetarian, \#veganism, \#cleaneating, and TikTok hashtags \#eatclean, \#healthyliving, \#detox, among others. Both spaces include ``body positivity'' cluster, with Twitter hashtags \#bodyconfidence, \#curvygirl, \#cute and TikTok hashtags \#curvemodel, \#fatacceptance. Twitter has an active ``diet'' community, which includes many related topics like \#ketoforbeginners, \#lowcarbs, etc. TikTok features several coherent topic clusters, including those about juicing and juice cleanse, the rapper Juicewrld, aquariums, and different ways to objectify a woman's body, like \#longlegs, \#hottie, \#curves, etc.

The biggest difference between the platforms is for topics related to eating disorders. Although hashtags like \#ana, \#anoreksja, \#edrecovery  were common to both platforms, TikTok also had recovery-related hashtags \#recoveryispossible and \#edrecovery that were tightly linked with body positivity topics as well as central \#foryoupage hashtags. Hashtags like  \#anarecovry, \#edawarewness, \#edsheeranrecoveryy, \#edrecocery, which were misspelled potentially to avoid TikTok content moderation algorithm, also linked to videos discussing recovery (see Table~\ref{tab:TikTok} in the Appendix). Unlike TikTok, Twitter contained explicit hashtags related to harmful pro-anorexia content like \#thinspo, a term used to glorify extreme thinness,  \#proana and \#promia, which refer to subcultures that glorify anorexia and bulimia, \#bonespo and \#deathspo for images of extreme thinness, etc. 
These hashtags linked to each other, as well as to hashtags related to self-harm and suicide glorifying subcultures (\#shtwt, \#goretwt), in a densely connected cul-de-sac of toxic topics. 

As on TikTok, eating disorders-related hashtags on Twitter were connected to popular mainstream topics ``diet,'' ``weightloss,'' and ``mentalhealth.'' This creates two concerns. First, vulnerable individuals seeking diet and weight loss advice may stumble on harmful hashtags that will lead them to content promoting anorexia. Second, recommendation algorithms may use these co-occurrences to expose users to ED-related content simply because they  viewed videos with hashtags ``diet'', ``weightloss.''

\subsection{Social Organization of Online Communities}
\label{sec:echo_chambers}
We explore the patterns of interactions between users posting about eating disorders and related topics on Twitter and Reddit. These interactions give us a view into the social organization of online communities.  

\subsubsection{Communities on Twitter}

\begin{figure}[ht]
    \centering
    \includegraphics[valign=c,width=0.6\linewidth]{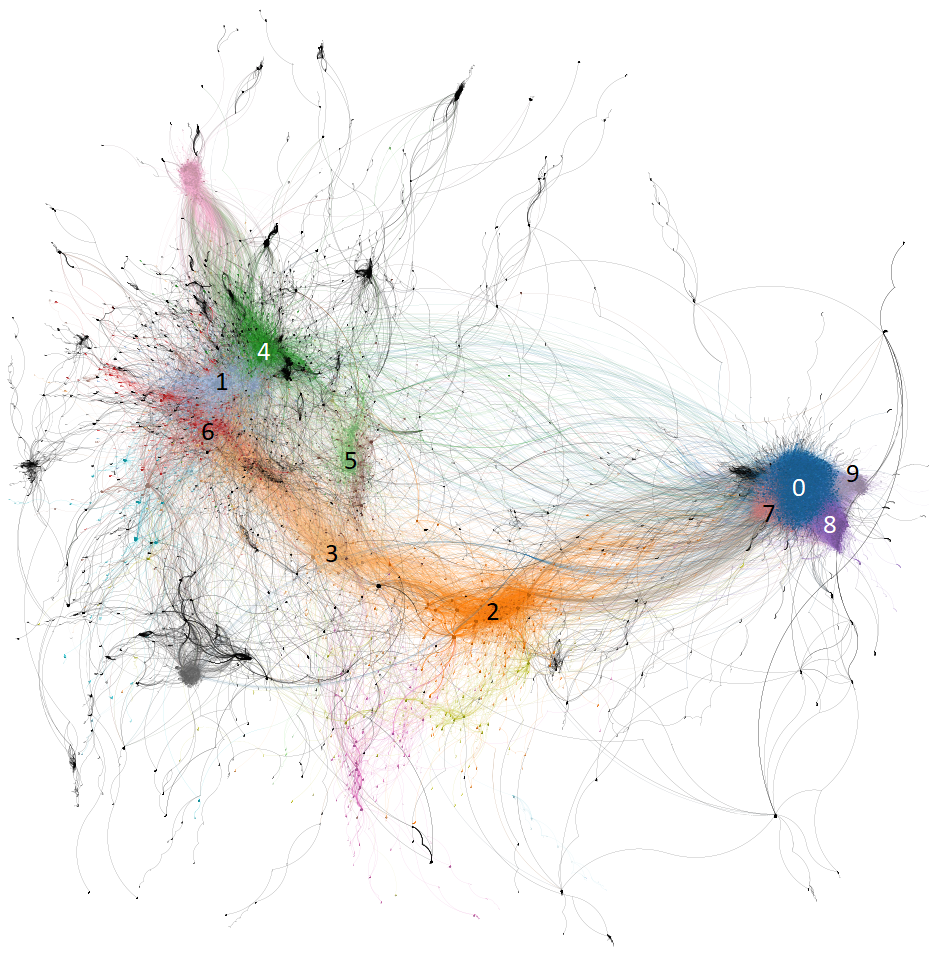}
    \includegraphics[valign=c,width=0.35\linewidth]{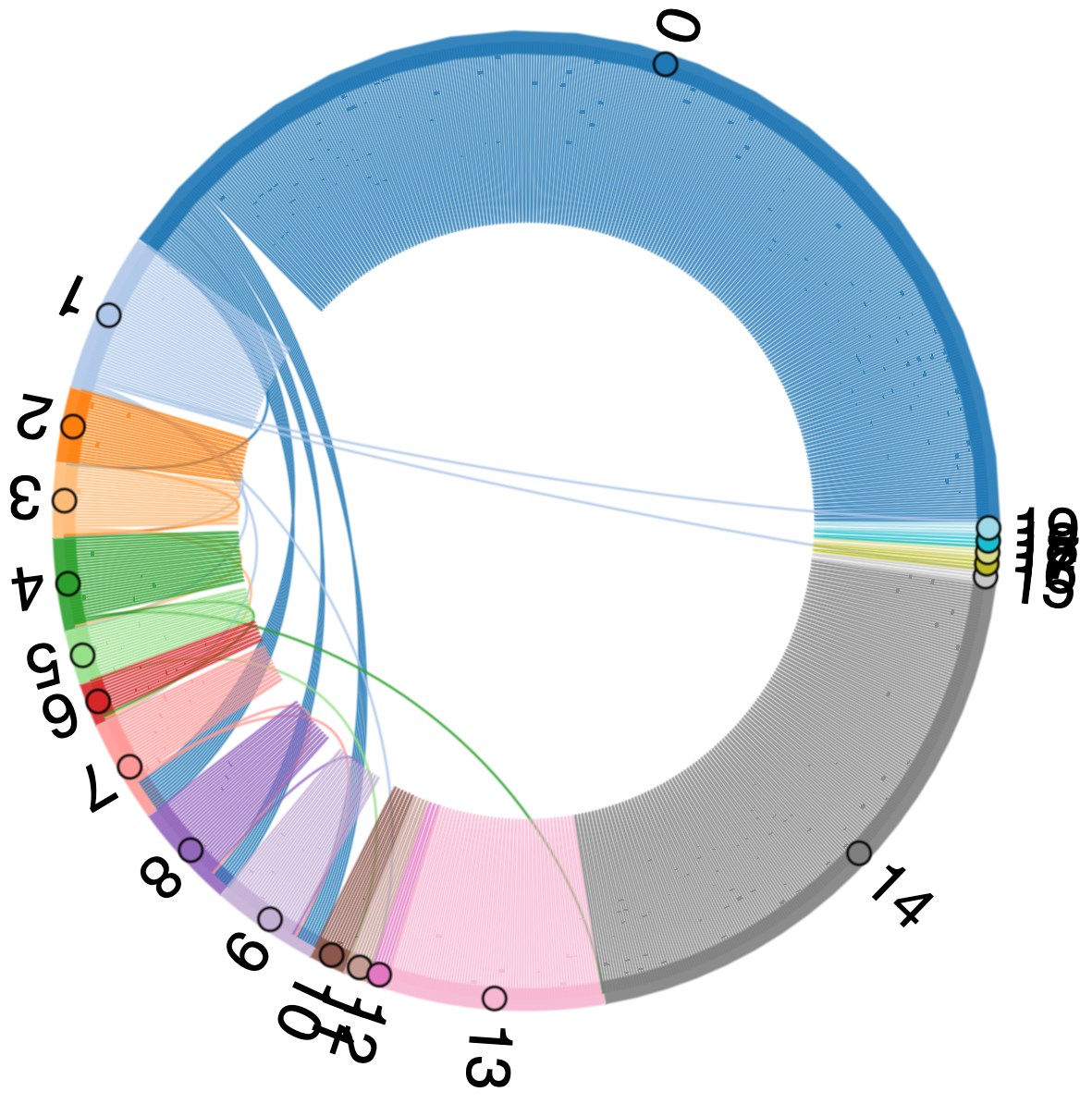}
\caption{Communities in the retweet network of Twitter. (Left) User network showing retweets between individual users. Colors correspond to different communities identified by the Louvain method. (Right) Chord diagram showing retweets within and between communities. The size of each chord represents the number of times members of a community retweeted themselves (self-retweeting), while the width of links shows the number of times they retweeted other communities.}
\label{fig:rt_network}
\end{figure}

As a proxy of attention, we construct a retweet network that links pairs of Twitter users who retweet each other at least once. 
Figure~\ref{fig:rt_network}(left) shows the retweet network. This network visualization is rendered using a force-directed layout \cite{atlas}, with nodes pushing away from each other while edges pull their connected nodes together. This spatial arrangement of users reflects the volume of retweets exchanged between them (users who retweet each other frequently are positioned closely to each other and vice-versa). Community detection using Louvain modularity~\cite{Blondel_2008} identifies 402 dense clusters---or communities---in this network. The largest 20 communities contain 71\% tweets and 40\% users in our dataset (Table \ref{tab:rt_comm_stats} in the Appendix). Due to differences in platform functionality and data collection procedures, we were not able to conduct a parallel analysis of communities on TikTok. 

The chord diagram in Fig.~\ref{fig:rt_network}(right)  shows community-level connections in the retweet network. The size of each chord represents the number of times members of that community retweeted each other (self-retweeting), while the width of links shows the number of times they retweeted another community. 
Most of the retweets are self-loops, suggesting that communities mostly pay attention to themselves, acting as echo chambers of like-minded users. Communities that frequently retweet each other, e.g., 0, 7, 8, and 9, are placed near each other in the network visualization.

\textbf{Profiling Communities.}
To get a sense of topics discussed within each community, we feed a random sample of 200 posts from each community to GPT-4 with the prompt: ``\emph{Given this list of posts, summarize the main ideas in 1 sentence}''. Varying random seeds during sampling does not substantively alter the summaries. We believe that the posts exhibit sufficient thematic consistency, ensuring the robustness of our random sampling approach.

The summaries of the ten largest communities are shown in Table~\ref{tab:comm_summaries} in the Appendix. 
Users in communities 0, 2, 7, 8, and 9 frequently use the term ``edtwt'' to self-identify as members of the eating disorders community on Twitter and promote eating disorders. For this reason, we refer to these communities as \textit{pro-anorexia} or \textit{pro-ana}. Interestingly, cluster 8 contains many Spanish language posts, and cluster 9 contains posts in Portuguese, suggesting these are international pro-anorexia communities. They are also placed tightly close to each other in Fig. \ref{fig:rt_network}(left). 
Although we do not have all the retweets made by the users but only those containing our search query terms, the patterns of connectivity that we observe suggest that they faithfully capture the structure of attention in the discussions of these topics. 
The sparseness of interactions between pro-anorexia  (0, 7, 8, 9) and other communities suggests that \textit{pro-anorexia communities on Twitter form an echo chamber}. 

Community 2, although also dedicated to eating disorder discussions, is well separated from the rest. This community takes a critical stance on eating disorders. For this reason, we refer to this community as \textit{anti-anorexia} or \textit{anti-ana}~\cite{oksanen2015pro}. 
This community uses counter-speech to engage with the pro-anorexia community, mostly to criticize it as being toxic, fatphobic, and harmful, but it interacts with other communities also. As a result, it is placed between pro-ana and other communities in the retweet network.

The remaining large communities are loosely connected, with no community truly isolated from the rest. These communities share posts about diet and weight loss but they are not as insular as the pro-ana community is. Community 1 discusses the risks and benefits of the \textit{keto diet}; communities 3 and 6 focus on issues surrounding the use of \textit{weight loss drugs} like Ozempic and Wegovy; community 4 looks at issues of \textit{healthy lifestyle and weight loss}, while community 5 covers \textit{body image} topics, like body positivity and self-acceptance. Communities 11, 12, and 14 are on other random issues irrelevant to eating disorders, and thus we exclude them from the analysis.

\subsubsection{Communities on Reddit}
Content on Reddit is organized around discussion forums, or subreddits, each focusing on a specific topic. However, users within one subreddit may mention other subreddits, which Reddit converts to a hyperlink. These hyperlinks connect discussion forums, providing a way for users to navigate between different topics and communities.

\begin{figure*}[tbh]
    \centering
    \includegraphics[width=0.7\linewidth]{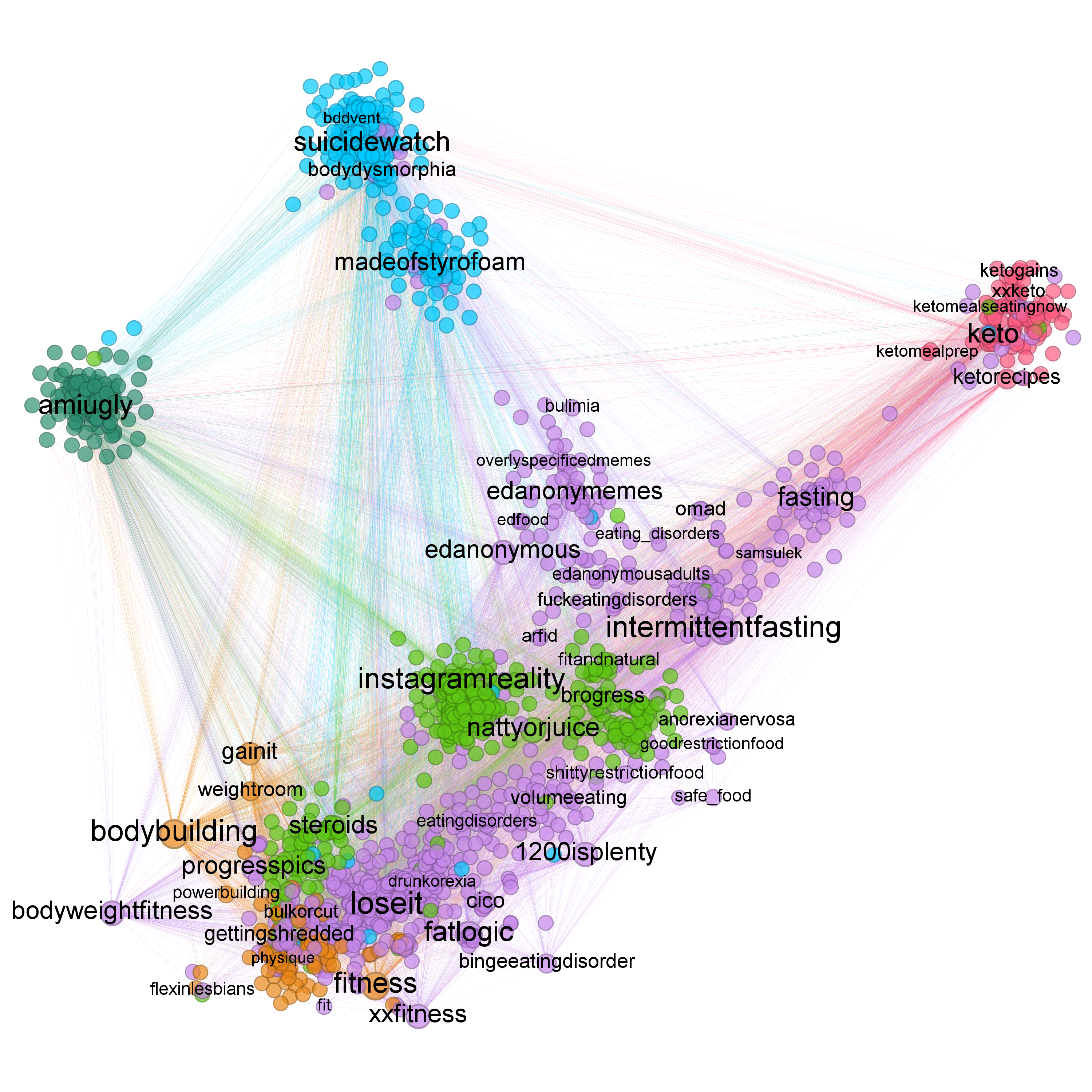}
  \includegraphics[width=0.8\linewidth]{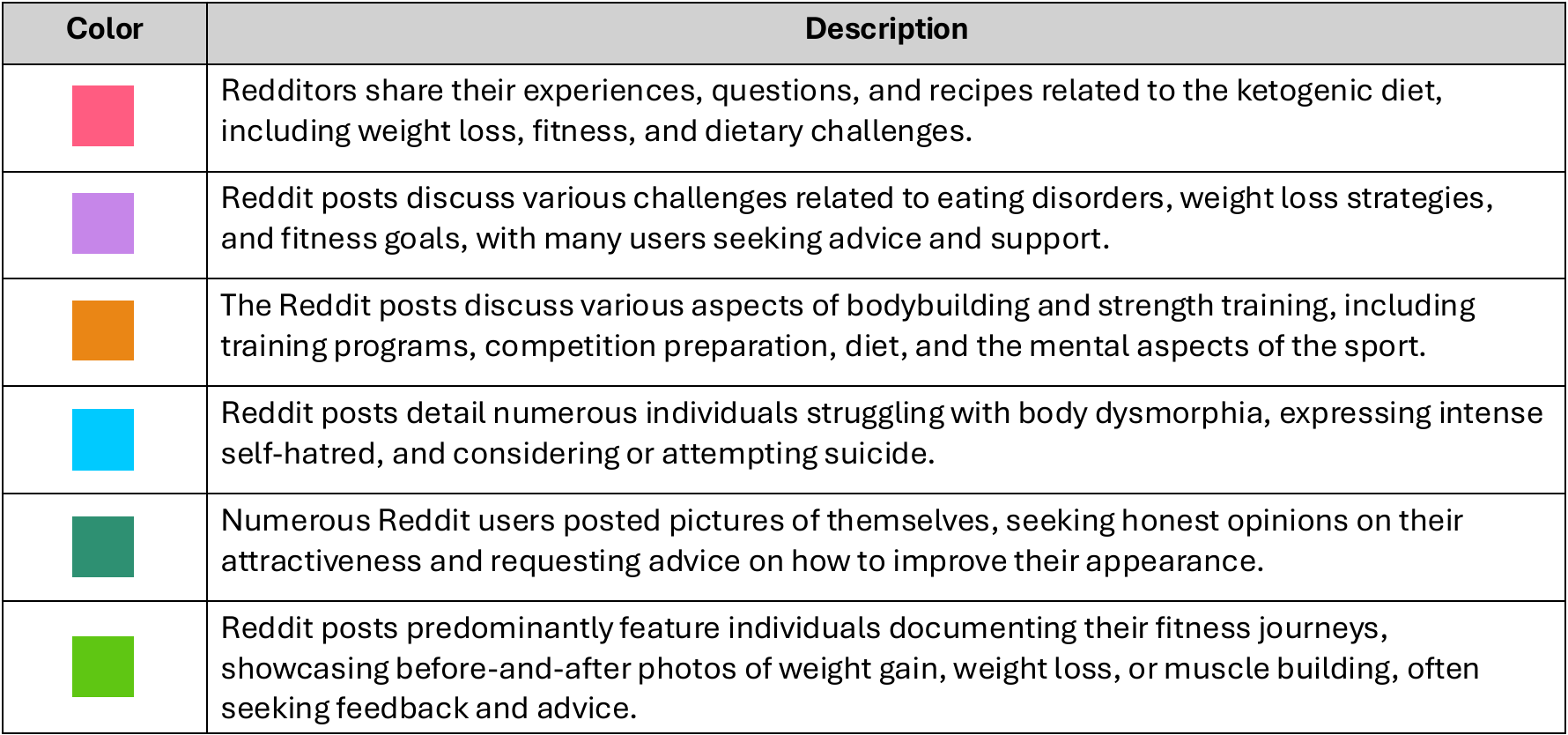} 
    \caption{Subreddit mention network.  Nodes represent subreddits, with node colors representing higher-level clusters. Links share the same color as the source subreddit, while node sizes are proportional to their degrees, indicating the level of interaction. 
    Description of each cluster provided by Gemini 1.5  summarizing 200 randomly sampled posts from each cluster.}
    \label{fig:reddit-mention}
\end{figure*}

To better capture interactions across subreddits, we used the original 54 unfiltered subreddits, along with their pre-sampled submissions and comments, to construct the subreddit mention network. Figure~\ref{fig:reddit-mention} visualizes the network, consisting of 1,950 subreddits with 18,202 mentions between them. Subreddits within the same higher-level cluster share a color to enhance visualization. Among these, 71 subreddits have been banned, 18 are gated, and 3 are quarantined. 

The mention network reveals distinct clusters, showing significant interaction across communities. Mental health-focused subreddits, such as \texttt{r/SuicideWatch}, \texttt{r/MadeOfStyrofoam} (self-harm), and \texttt{r/BodyDysmorphia}, are closely linked at the top and marked in light blue. These mental health subreddits are also connected to clusters like the keto community (dark pink) and the body image cluster (dark green).

The lower portion of the network contains a dense cluster of interrelated subreddits that discuss topics ranging from weight loss (\texttt{r/loseit}) and bodybuilding (\texttt{r/bodybuilding}) to eating disorders like anorexia (\texttt{r/anorexianervosa}) and bigorexia (\texttt{r/GettingShredded}). 

Although we lack data for all mentioned subreddits, the observed connectivity patterns likely reflect the attention structure within these discussions.
The structure reveals strong connections and awareness between weight management, fitness, restrictive diet forums, and mental health spaces. Unlike Twitter, eating disorders-related subreddits are also well integrated within the social organization of forums on Reddit.

\subsection{User Engagement in Online Communities}

Online communities provide emotional support, creating a safe space to vent, i.e., express negative emotions, and receive support in the form of positive emotions like  love~\cite{McCormack2010}. We analyze patterns of emotional engagement across platforms and show they are consistent with emotional support. This dynamic is a key component of the group processes that keep members engaged in online communities, even those that are harmful to them.

\begin{figure*}[tbh]
    \centering
    \includegraphics[width=0.95\linewidth]{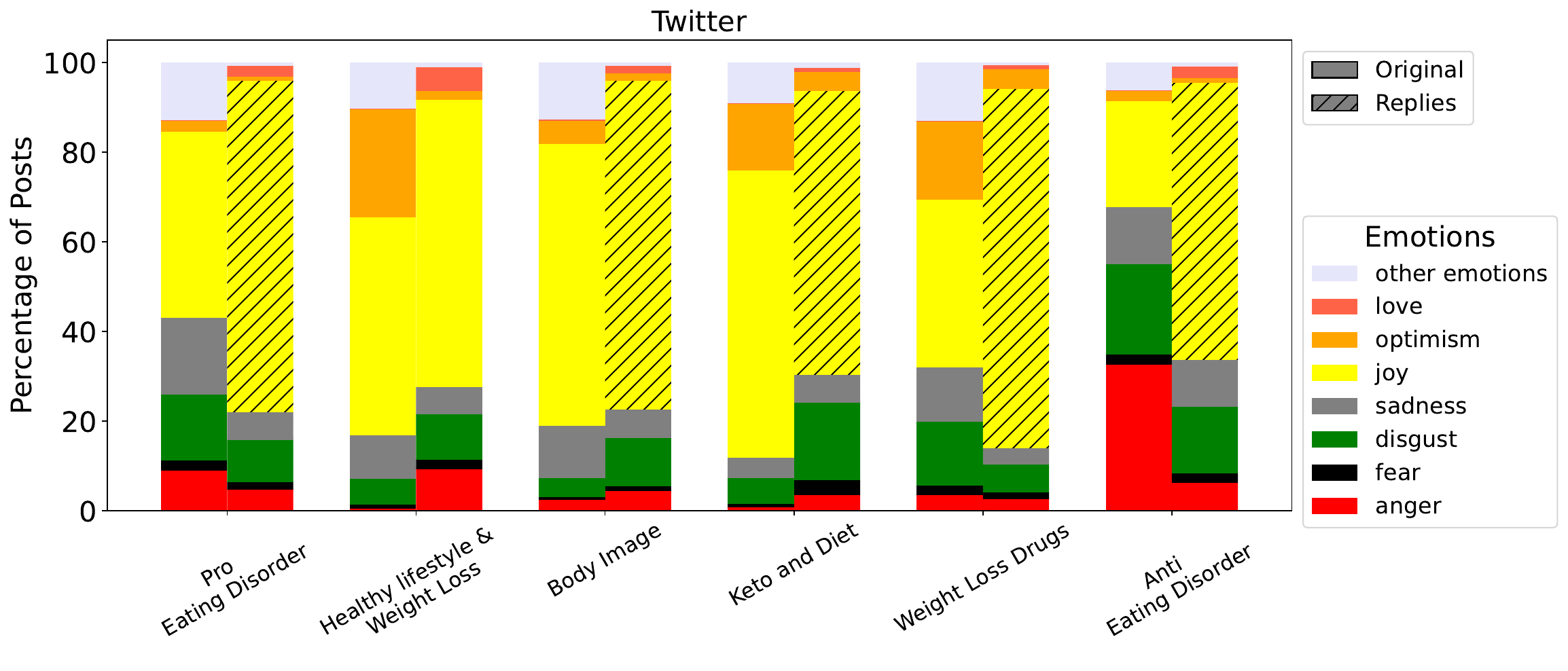}
    \includegraphics[width=0.95\linewidth]{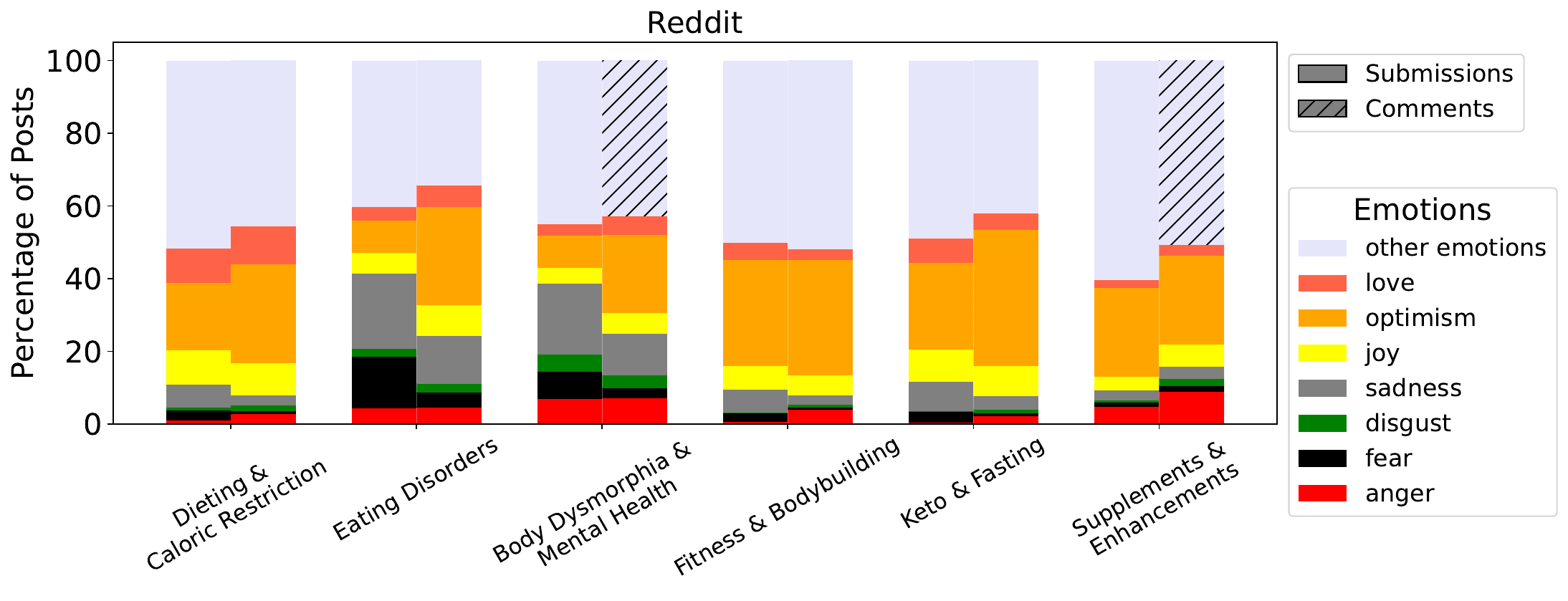}

    \caption{
    Community emotion analysis. For each  (a) community on Twitter, (b) Reddit discussion forum, or (c) TikTok hashtag, we aggregate the emotion confidence scores across all \textbf{original} tweets/submissions (open bars) and replies/comments (hashed bars) and show the share of posts with each emotion.
    }
    \label{fig:emo}
\end{figure*}

\begin{figure*}[tbh]
    \centering
    \includegraphics[width=0.95\linewidth]{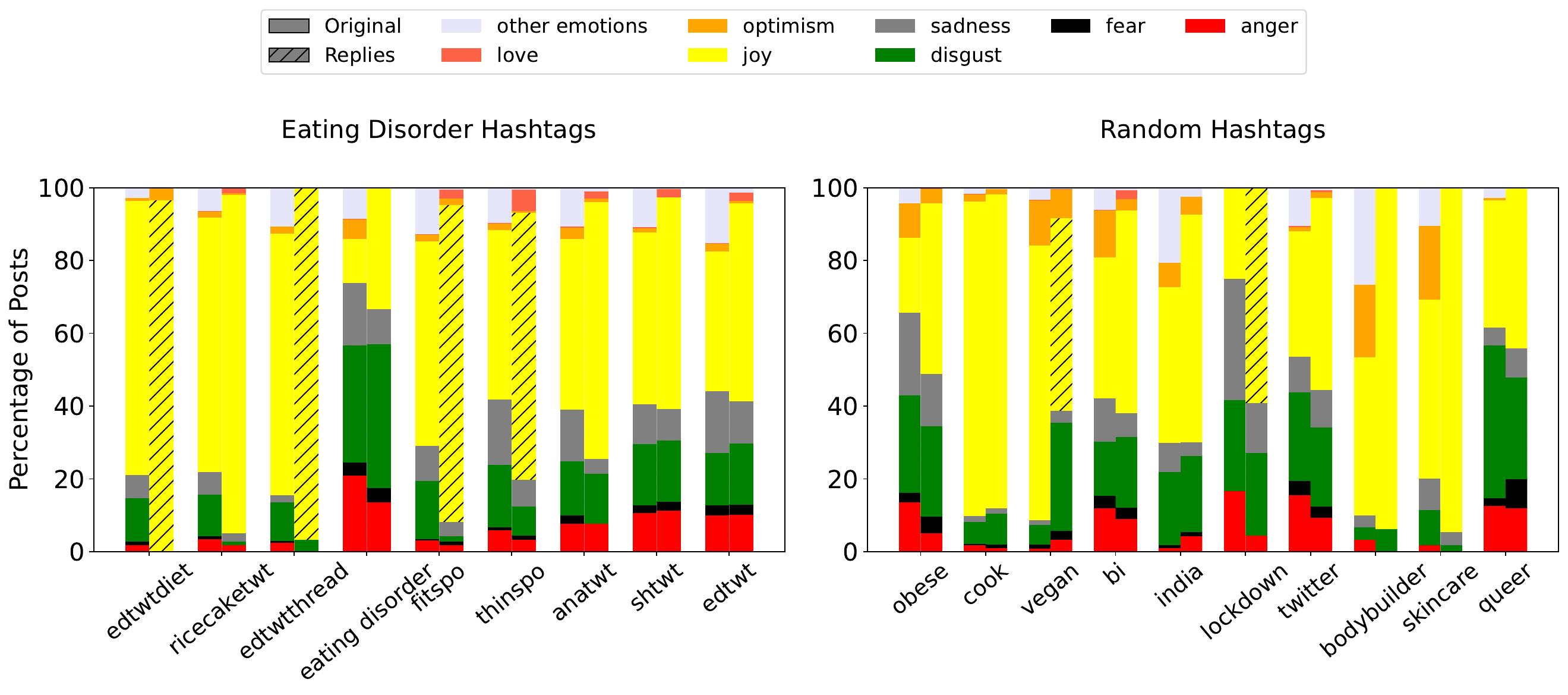}
    \includegraphics[width=0.95\linewidth]{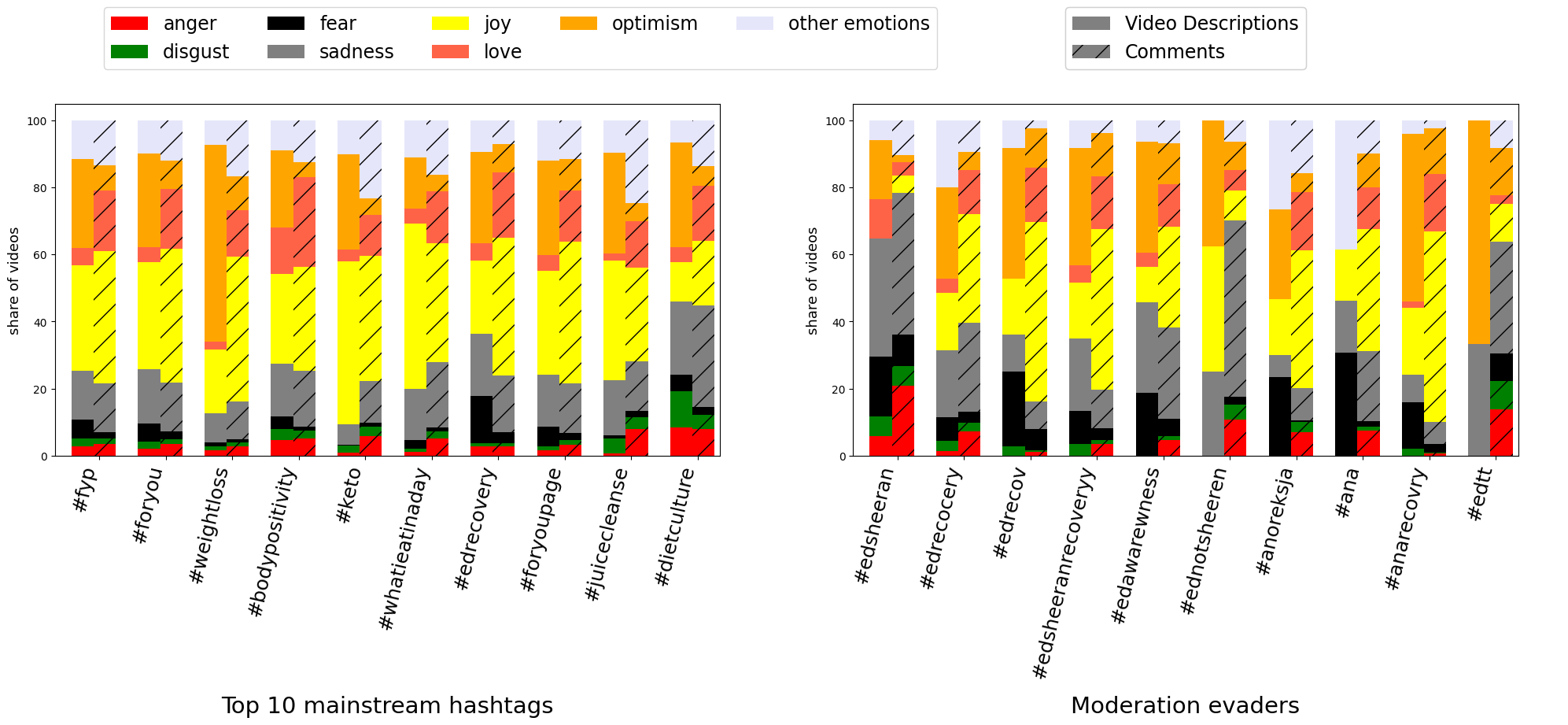}
    \caption{
    Emotion analysis of user engagement. The share of posts and comments with a specific hashtag that expresses an emotion. The posts are (a) tweets on Twitter and (b) video descriptions on TikTok. Posts related to eating disorder hashtags or moderation evaders on TikTok generally express more negative emotions, consistent with venting.  
    }
    \label{fig:emo-hashtag}
\end{figure*}

\subsubsection{Emotions in Online Communities}
The emotion detection models return confidence scores for anger, disgust, fear, sadness, joy, love, optimism, and trust. 
Figure~\ref{fig:emo} compares the share of posts expressing an emotion among original posts and replies on Twitter and Reddit. Posts are grouped by related communities on Twitter (top) and subreddits on Reddit (bottom). 
Twitter is more emotional, with a higher share of posts (original tweets and submissions) and replies containing emotions. It is also the most joyful, although joy and optimism are also highly prevalent on Reddit as well. Replies (and comments) are generally more emotional than original posts (and submissions). However, there are systematic differences in the distribution of emotions across communities. On Twitter, the eating disorder communities, both pro-anorexia and anti-anorexia, express more negative emotions (\textit{sadness}, \textit{disgust}, \textit{fear}, \textit{anger}) in posts,  compared to other communities, while replies (hashed bars) express more positive emotions like \textit{joy} and \textit{love}, consistent with communities providing emotional support. The community discussing weight loss drugs is the only other community to show similar patterns of emotional support. 
Reddit shows similar trends (Fig.~\ref{fig:emo}bottom), where forums discussing eating disorders and mental health exhibit more negative emotions in submissions, but higher levels of positive emotions in comments to the submissions, characteristic of online communities providing emotional support.

We cannot evaluate emotions at a community level on TikTok: instead, we compare them across hashtags. 
Figure~\ref{fig:emo-hashtag} shows the distribution of emotions in posts tagged with a specific hashtag on Twitter and TikTok. We separate hashtags related to eating disorders from random hashtags on Twitter, and moderation evaders from mainstream hashtags on TikTok since videos tagged with moderation evading hashtags discuss eating disorders topics (see  Table~\ref{tab:TikTok} in the Appendix). Twitter and TikTok differ in the emotional tone: there is more \textit{love} and \textit{optimism} on TikTok and more \textit{disgust} and \textit{joy} on Twitter. On both platforms, however, posts with hashtags related to eating disorders tend to be more negative overall, expressing more  (\textit{anger}, \textit{fear}, \textit{sadness}) compared to random (mainstream) hashtags.


\begin{figure*}[ph]
    \centering
    \begin{tabular}{c}
    \includegraphics[width=0.8\linewidth]{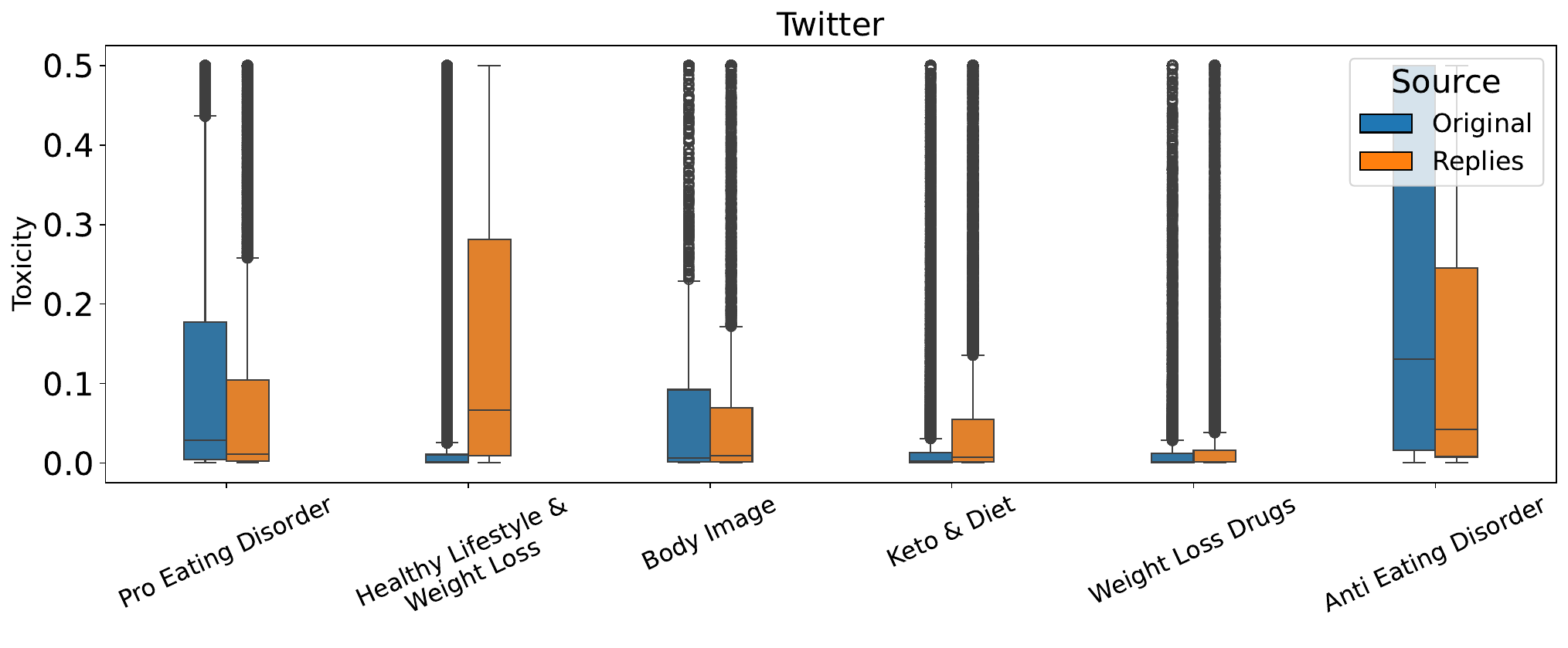} \\
    (a) \\
    \includegraphics[width=0.8\linewidth]{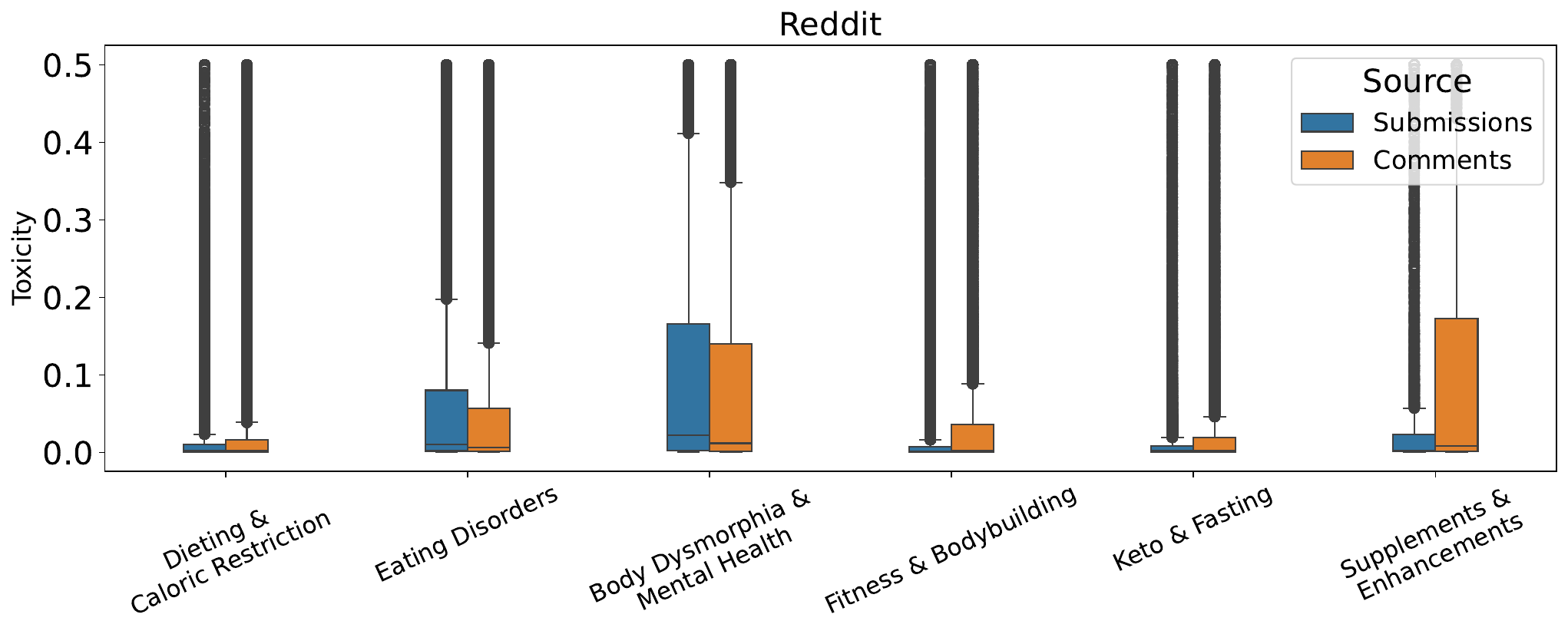} \\ (b) \\    
    \includegraphics[width=0.8\linewidth]{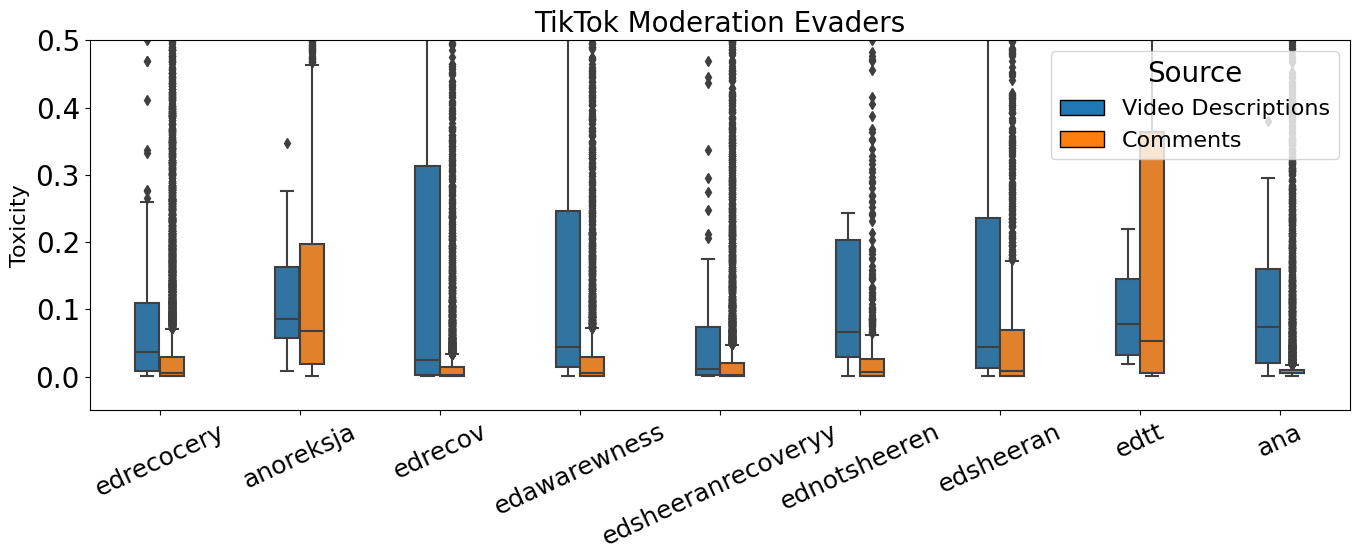}
    \\ (c) \\
    \includegraphics[width=0.8\linewidth]{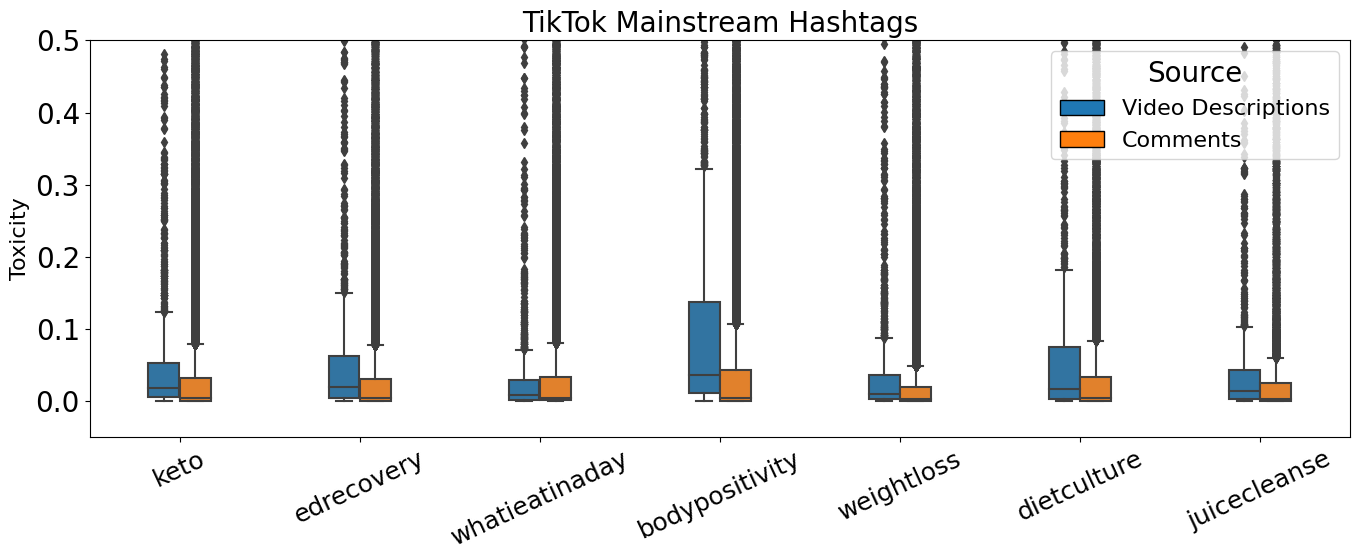}
    \\ (d)
        \end{tabular}
    \caption{Toxicity analysis of online communities. Boxplot shows the distribution of toxicity scores for posts and comments for online communities on (a) Twitter, (b) Reddit, (c) moderation evaders and (d) mainstream hashtags on TikTok.
    }
    \label{fig:toxicity}
\end{figure*}

\subsubsection{Toxicity in Online Communities}
One of the most important questions is how much harmful content promoting eating disorders online communities create. However, identifying harmful content is challenging: what constitutes harmful material can be highly contextual, subjective, and ambiguous. Pro-anorexia communities continually evolve their tactics to evade detection,  using coded language or passing off material as wellness content. 
Given these challenges, we focus instead on the use of toxic language. 
Toxicity has been shown to negatively impact the life-cycle of online communities~\cite{mohan2017impact} and degrade the emotional tone of discussions~\cite{chang2023feedback}. Toxicity can be especially problematic in mental health communities, where it can exacerbate the stigma surrounding health conditions. Toxicity detection (see Methods) serves as a valuable tool for identifying concerning trends, guiding content moderation efforts, and assessing the overall health of online discussions.

Figure~\ref{fig:toxicity} presents the distribution of toxicity scores in posts and replies grouped by communities on Twitter (Fig.~\ref{fig:toxicity}(a)) and Reddit (Fig.~\ref{fig:toxicity}(b)), and hashtags on TikTok (Fig.~\ref{fig:toxicity}(c)).
While there is a wide variance of toxicity, overall, TikTok users who post content under moderation evading hashtags express higher levels of toxicity than others. Consistently, TikTok videos exhibit more toxicity than what shows up in their comments. 

The eating disorder and mental health communities on both Twitter and Reddit share similarities: posts tend to be more toxic than comments, and these communities are generally more toxic than others. This is consistent with users in these forums frequently seeking emotional support through self-disclosure, which often involves explicit language (e.g., obscenities), and for members of marginalized communities, using reclaimed slurs. As a result, toxicity detection algorithms are more likely to mistakenly flag these posts as toxic \cite{garg2023handling,dorn2024harmful}. 

Notably, posts from members of the anti-anorexia community on Twitter, which frequently criticize harmful practices of the pro-anorexia community, are often flagged as toxic due to the use of profanity and insults. However, these narratives, which can be considered counter-speech that actively challenge unhealthy norms and behaviors, pose minimal risk to the community despite being labeled as highly toxic.



\subsection{Discussion}

Many theoretical lenses have been used to study online spaces devoted to eating disorders~\cite{pater2017defining,choukas2022perfectstorm}; however, they fail to account for group dynamics. Our paper describes interactions within online communities as a social dynamic in which communities provide information as well as social and emotional support. On Twitter, where moderation is lax, these dynamics trap individuals in toxic echo chambers that not only expose them to pro-eating disorder content but also isolate them from recovery-oriented perspectives. Participants then contribute or amplify more pro-anorexia content, perpetuating the cycle and ensnaring new members.

This social dynamic is similar to \textit{online radicalization}, a process by which individuals adopt ever more extreme beliefs that lead them to deviant behaviors, such as violence and terrorism. 
The 3N model of radicalization explains~\cite{Kruglanski2014} how the loss of personal significance (\textit{Need} for significance) pushes individuals to create \textit{Narratives} justifying extreme behaviors that allow them to regain significance. These narratives are validated by  \textit{Networks}, or communities, that connect individuals to like-minded others while isolating them from alternative worldviews.
In the context of eating disorders, loss of personal significance may be triggered by negative body image, despair, or bullying that drive vulnerable youth to seek weight loss advice online~\cite{Ging2018,choukas2022perfectstorm,pater2016hunger}. Pro-anorexia communities fulfill that \textit{need for significance} by allowing individuals to set \textit{body goals} (e.g., ``lose 20 lbs before prom'') and track progress for \textit{accountability} (both common hashtags in our data). 
Communities also create a collective identity~\cite{polletta2001collective} by having members self-label themselves with hashtags that express solidarity with a chosen  group~\cite{barron2022quantifying}, e.g., ``edtwt'' or ``ricecaketwt''.
Once individuals join a pro-anorexia community, or radical \textit{network},  group dynamic processes keep them engaged~\cite{Kruglanski2014}. This includes providing emotional support and validation.

Discussions within pro-anorexia communities create harmful \textit{narratives} that promote extreme behaviors. 
Twitter, which was the only platform in our study without robust moderation, narratives included explicitly harmful pro-anorexia content.
Our analysis shows that content moderation can disrupt these processes. As a result of moderation on TikTok and Reddit, though eating disorders-related discussions exist, they are more recovery-oriented and better integrated within mainstream discussions.


\subsubsection{Limitations}
Some methodological considerations may affect the conclusions of our study. Our analysis of TikTok data focused on textual video descriptions provided by the creators and did not examine video content.

We collected Twitter and TikTok data by searching for content with specific keywords. While this method samples the content that is generated by users on these platforms, it does not sample the content that users see. Feed recommendation algorithms may selectively present more harmful content to users. A recent study audited TikTok user social feeds and showed that TikTok recommended dramatically more toxic ED videos to people struggling with EDs than control subjects~\cite{griffiths2024does}.

We used two different emotion models---SpanEmo-based model for Twitter and TikTok and a GoEmotions-based model for Reddit---which could raise concerns about standardization and potential biases inherent to each model for a robust cross-platform analysis. However, this approach is intentional and justified: SpanEmo is a state-of-the-art model for detecting a range of emotions in social media text~\cite{chochlakis2023emotion},  while the GoEmotions model was trained specifically on Reddit data offering domain-specific accuracy tailored to that platform. Importantly, both models recognize a similar range of emotions. 

Importantly, our study cannot discriminate between individuals with eating disorders and those who are merely discussing disordered behaviors and beliefs. We also do not have evidence that joining pro-anorexia conversations results in people developing eating disorders or exacerbating an existing illness. We argue, however, that such communities are harmful in that they normalize and promote disordered eating behaviors, potentially delaying recovery for individuals struggling with anorexia and bulimia.

\subsubsection{Ethics Statement}
This research addresses sensitive mental health topics, requiring careful consideration of ethical implications. Although we relied solely on publicly available data, in compliance with platform terms of service and an IRB-reviewed study protocol, we took additional precautions to protect privacy. These included removing identifiable information and focusing on aggregated data analysis. To safeguard the well-being of our research team, we avoided collecting images and conducted regular team check-ins.

\section{Conclusion}
Our study investigates the social dynamics of conversations about eating disorders across three social media platforms: Twitter, Reddit, and TikTok. Using a combination of network and language analysis, we characterize these conversations to reveal their semantic and social organization. We find that while online communities can provide social connections, a shared group identity, and emotional support, they also risk becoming toxic echo chambers that normalize and amplify harmful pro-anorexia content. However, effective content moderation significantly reduces these harmful social dynamics.

To the best of our knowledge, this is the first study to apply this specific mechanism and model to understanding mental health conditions like eating disorders. Our work emphasizes the critical role of social dynamics in explaining mental health trends and adds to the growing body of research linking social media to behaviors that negatively impact body image and mental well-being. The social dynamic mechanisms we explore offer a framework for understanding these phenomena, linking them to other harmful behaviors, and suggesting potential mitigation strategies.

As evidence grows of social media's role in propagating psychological distress~\cite{bartholomew2012mass}, understanding the interplay between individual psychological states and online communities is essential. Social media platforms actively shape user behaviors through algorithms and social dynamics, amplifying both positive and negative experiences. 
By investigating this dynamic, researchers can identify the specific ways in which social media environments contribute to or mitigate psychological distress. Such insights are essential for developing content moderation policies that leverage online interactions to foster resilience and recovery. 

\section{List of abbreviations}
\begin{itemize}
    \item ED -- Eating Disorders
    \item GPT-4 -- Generative Pre-trained Transformer 4
    \item API -- Application Programming Interface

\end{itemize}

\section{Appendix}
\subsection{Data Collection}
\begin{table}[ht]
\small
\caption{Search terms  used to retrieve ED posts from Twitter and TikTok.}
\begin{tabular}{p{0.9\linewidth}}
\toprule
\multicolumn{1}{c}{\textbf{Keywords}}             \\ \hline
\emph{thinspo}, \emph{edtwt}, \emph{proana}, \emph{proanatips}, \emph{anatips}, \emph{meanspo}, \emph{fearfood}, \emph{sweetspo}, \emph{eatingdisorder}, \emph{bonespo}, \emph{promia}, \emph{redbracetpro}, \emph{bonespo}, \emph{m34nspo}, \emph{fatspo}, \emph{lowcalrestriction}, \emph{edvent}, \emph{WhatIEatInADay}, \emph{Iwillbeskinny}, \emph{thinspoa}, \emph{ketodiet}, \emph{skinnycheck}, \emph{thighgapworkout}, \emph{bodyimage}, \emph{bodygoals}, \emph{weightloss}, \emph{skinnydiet}, \emph{chloetingchallange}, \emph{fatacceptance}, \emph{midriff}, \emph{foodistheenemy}, \emph{cleanvegan}, \emph{keto}, \emph{cleaneating}, \emph{intermittentfasting}, \emph{juicecleanse}, \emph{watercleanse}, \emph{EDrecovery}, \emph{bodypositivity}, \emph{dietculture}, \emph{slimmingworld}, \emph{losingweight}, \emph{weightlossmotivation}, \emph{healthyliving}, \emph{weightlosstips}, \emph{weightlossjourney}, \emph{wegovy}, \emph{semaglutide}, \emph{ozempic}.
                                             \\ \bottomrule
\end{tabular}
\label{app:search-terms}
\end{table}



\begin{table}[ht]
\small
\caption{Subreddits collected in our data. The prefix "r/" was removed for clarity.}
\begin{tabular}{p{0.9\linewidth}}
\toprule
\multicolumn{1}{c}{\textbf{Subreddit (r/)}}             \\ \hline
steroids, Brogress,  BulkOrCut, GettingShredded, weightroom, nattyorjuice, powerbuilding,  bodybuilding, gainit, bodyweightfitness, Instagramreality, ketorecipes, ShittyRestrictionFood, progresspics, goodrestrictionfood, EDanonymemes, Volumeeating, amiugly, safe\_food, FlexinLesbians, xxfitness, xxketo, EDAnonymous, BingeEatingDisorder, ARFID, EdAnonymousAdults, EatingDisorders, fuckeatingdisorders, eating\_disorders, bulimia, AnorexiaNervosa, BodyDysmorphia, BDDvent, fit, ketogains, fasting, omad, 1200isplenty, CICO, intermittentfasting, loseit, Fitness, keto, MadeOfStyrofoam, drunkorexia, SuicideWatch                                                                                           \\ \bottomrule
\end{tabular}
\label{tab:subreddits}
\end{table}

\subsection{Community Profiles}
\label{app:gpt}


\begin{table}[ht]
\footnotesize
    \centering
\addtolength{\tabcolsep}{-3.3pt}
\begin{tabular}{lccccccccccc}
\hline
\textbf{Comm}      & \textbf{0} & \textbf{1} & \textbf{2} & \textbf{3} & \textbf{4} & \textbf{5} & \textbf{6} & \textbf{7} & \textbf{8} & \textbf{9} &  \\ \hline
\# of users & 61,954     & 24,400     & 21,887     & 20,631     & 9,901      & 9,031      & 9,000      & 8,084      & 7,702      & 7,020      &            \\
\# of tweets       & 805,249    & 112,674    & 32,883     & 37,788     & 193,348    & 24,395     & 21,369     & 82,702     & 70,764     & 71,970     &         \\ \hline
\textbf{Comm}      & \textbf{10} & \textbf{11} & \textbf{12} & \textbf{13} & \textbf{14} & \textbf{15} & \textbf{16} & \textbf{17} & \textbf{18} & \textbf{19} & \textbf{total} \\ \hline
\# of users &   6,477  & 6,158     &  5,181    &  4,528   &   3,682    &  3,672    &  3,360     &  3,163     &  3,086    &   2,865    &  221,887         \\
\# of tweets       &  15,796   & 9,254   & 7,019  &  103,177   &  260,971   &  5,338   & 4,881  &  5,065   &   4,612   &   7,021   & 1,876,276        \\ \hline
\end{tabular}
\addtolength{\tabcolsep}{3.3pt}
\caption{Number of users (community size) and tweets in the top 20 largest communities respectively and in total.}
\label{tab:rt_comm_stats}
\end{table}
\begin{table}[ht]
    \centering
    \footnotesize
    \begin{tabular}{@{}c|p{0.65\linewidth}|p{0.1\linewidth}||c|p{0.1\linewidth}@{}}
id & Summary of community posts & Label & id & Label\\ \hline
 {0} &  The tweets revolve around the online eating disorder community (edtwt), sharing tips, thinspo (thin inspiration), meanspo (mean inspiration), fasting strategies, and discussing body image and weight loss goals, often in a way that promotes disordered eating behaviors.
& Eating Disorder
& 10 & Body Image\\
\textcolor{gray}{1} & \textcolor{gray}{The tweets cover a range of topics related to ketogenic diets, weight loss, metabolic health, and low-carb recipes, with discussions on the effectiveness of keto for various health conditions, debates on prescribing obesity drugs to children, and personal testimonials about the benefits of a keto lifestyle} 
& \textcolor{gray}{Keto \& Diet}
& \textcolor{gray}{11} & \textcolor{gray}{Spam} 
\\
{2} & The tweets express strong negative sentiments towards "edtwt" (presumably "eating disorder Twitter"), criticizing it for being toxic, fatphobic, and harmful, with calls to abolish it and stop interacting with its content
& Anti-Eating Disorder 
& 12 & Spam
 \\
\textcolor{gray}{3} & \textcolor{gray}{The tweets discuss the controversial use of the diabetes drug Ozempic for weight loss, the impact of its shortage on diabetic patients, the cost of the medication, and related topics such as body positivity, keto diets, and the role of influencers and celebrities in promoting certain health trends and products} 
& \textcolor{gray}{Weight Loss Drugs} 
& \textcolor{gray}{13} & \textcolor{gray}{Healthy lifestyle \& Weight Loss}
\\ 
{4} & The tweets cover a variety of health and wellness topics, including weight loss methods, dietary plans, fitness advice, healthy eating, keto diet, fasting, moxibustion, and motivational messages for maintaining a healthy lifestyle 
& Healthy lifestyle \& Weight Loss 
& 14 & Spam
 \\
\textcolor{gray}{5} & \textcolor{gray}{The tweets cover a variety of personal updates, including fitness activities, body positivity, nudism, modeling, and social interactions, with some tweets promoting content or expressing motivational thoughts} 
& \textcolor{gray}{Body Image} 
& \textcolor{gray}{15} & \textcolor{gray}{Keto \& Diet}
\\
6 & The tweets discuss various topics related to weight loss, dieting, and health, including the benefits of intermittent fasting, ketogenic diets, and the use of medications like Semaglutide for obesity and diabetes management, while also mentioning challenges such as supply issues for weight loss drugs and the importance of exercise and healthy eating habits 
& Weight Loss Drugs 
& 16 & Keto \& Diet
\\
\textcolor{gray}{7} &  \textcolor{gray}{The tweets are primarily from individuals within the eating disorder Twitter community (edtwt) sharing personal updates, seeking mutuals (moots), discussing weight goals, and exchanging tips and thinspiration, with some also mentioning other interests such as music and anime} 
& \textcolor{gray}{Eating Disorder} 
& \textcolor{gray}{17} & \textcolor{gray}{Healthy lifestyle \& Weight Loss}
\\
8 & The tweets revolve around the online community known as "edtwt" (eating disorder Twitter), discussing topics such as bulimia, thinspiration (thinspo), seeking mutual followers within the community, meanspo (mean inspiration), bonespo (bone inspiration), weight loss, pro-anorexia (proana) sentiments, and sharing tips and experiences related to eating disorders 
& Eating Disorder
& 18 & Keto \& Diet 
\\
\textcolor{gray}{9} &  \textcolor{gray}{The tweets revolve around discussions and content related to eating disorders, specifically within the "edtwt" community, with a focus on promoting extreme thinness (thinspo, bonespo), sharing meanspo (mean-spirited motivation), and offering tips and support for disordered eating behaviors} 
& \textcolor{gray}{Eating Disorder}
& \textcolor{gray}{19} & \textcolor{gray}{Weight Loss Drugs}
\\
\hline
    \end{tabular}
    \caption{Twitter community profiles. GPT-4-generated summaries of posts in the ten largest communities in the Twitter retweet network. The table also shows the labels we assigned to the twenty largest communities based on their summaries.
    }
    \label{tab:comm_summaries}
\end{table}

\begin{table}[ht]
\footnotesize
    \centering
\addtolength{\tabcolsep}{-3.3pt}
\begin{tabular}{p{0.13\linewidth}|p{0.865\linewidth}}
\textit{Hashtag} & \textit{GPT-4 Summary of video descriptions} \\ \hline 
 edsheeran & The video content centers on eating disorder recovery and mental health awareness, featuring personal narratives of recovery journeys, coping strategies, and support systems. Common themes include self-acceptance, family support, and the complex relationship with food (including discussions of ``safe'' and ``fear'' foods). Content creators often use humor as a coping mechanism while emphasizing the importance of seeking professional help and maintaining supportive relationships.
 \\ \hline
\textcolor{gray}{edrecocery} & \textcolor{gray}{
The video content primarily focuses on eating disorder recovery journeys, emphasizing support networks and personal growth. Key themes include confronting fear foods, promoting mental health awareness, and cultivating self-acceptance. Content creators share both challenges and victories, particularly around food-related anxieties and holiday celebrations, while encouraging community support and self-compassion throughout the recovery process.
} \\ \hline
edrecov & 
The videos primarily focus on eating disorder recovery journeys, using hashtags like \#edrecov and \#anarecovery to document both challenges and victories. Key themes include mental health awareness, overcoming food-related anxieties (\#fearfoodchallenge), and building supportive communities (\#strongertogether). Content emphasizes personal growth through recovery milestones (\#recoverywins), managing setbacks (\#relapsehappens), and encouraging self-acceptance and empowerment (\#breakfree, \#stopcaloriecounting).
\\\hline
 \textcolor{gray}{edshee\-ran\-recoveryy} & \textcolor{gray}{
 These video descriptions highlight themes of eating disorder recovery, mental health awareness, and self-care. They feature hashtags like \#edrecovery and \#mentalhealthmatters, with individuals sharing personal struggles, recovery milestones, and community support. Key topics include fear of food, body image challenges, and self-love, reflecting a collective journey of resilience and growth in overcoming mental health obstacles.
 } \\ \hline
edawa\-re\-wness & 
These video descriptions cover themes of eating disorder recovery, mental health awareness, and personal growth. They highlight challenges, celebrate recovery progress, and emphasize support and understanding for those struggling. Topics include anxiety, depression, body image, and self-acceptance, promoting self-worth beyond appearance. The community aspect fosters solidarity, encouraging individuals to seek help, share experiences, and reduce stigma. Overall, the descriptions create a supportive space for connection and advocacy in mental health and eating disorder awareness.
\\ \hline
\textcolor{gray}{ednotshee\-ran} & \textcolor{gray}{
These video descriptions explore eating disorders (ED) and mental health awareness, focusing on personal recovery journeys, relapses, and the serious nature of EDs. Themes include challenging stereotypes, addressing body image and dysmorphia, and confronting societal beauty standards. Support and empowerment are central, offering encouragement, affirmations, and creative expression through platforms like TikTok. Awareness and advocacy are emphasized, aiming to destigmatize EDs and promote mental health support through shared experiences and personal insights.
}\\ \hline
anoreksja & 
These video descriptions focus on eating disorders (anorexia, bulimia), recovery, self-care, and body positivity. They share personal recovery journeys, self-love tips, gratitude for followers, and encouragement to fight against EDs. Content includes food choices, recipes, and seeking support, aiming to raise ED awareness, support recovery, and promote a positive mindset toward food and body image.
\\ \hline
\textcolor{gray}{ana} & \textcolor{gray}{
These video descriptions delve into themes of eating disorder (ED) recovery, mental health, and food relationships. Key focuses include the desire for a healthier connection with food, personal recovery journeys with hashtags like \#edrecovery, and discussions on safe and fear foods. Mental health awareness covers anxiety, depression, and body dysmorphia, stressing the importance of support. Recovery challenges and triumphs are highlighted, along with a caution against glamorizing EDs. Expressions of longing for normalcy in food relationships reveal a complex interplay between EDs, recovery, mental health, and societal views on food and body image.
}\\ \hline
anarecovry & 
These video descriptions highlight eating disorder recovery, self-acceptance, and confronting fear foods. They focus on personal victories and challenges in recovering from anorexia and bulimia, especially around facing fear foods. Themes of self-acceptance, mental health awareness, and empowerment are central, the promotion of autonomy, enjoyment of food without guilt, and seeking help. Support from friends, family, and online communities is emphasized, along with messages of positivity and motivation to inspire viewers on their healing journey.
\\ \hline
\textcolor{gray}{edtt} & \textcolor{gray}{
These video descriptions focus on mental health, particularly eating disorders, highlighting personal struggles with body image, control, and dieting. Recovery and seeking inspiration are central, alongside expressions of frustration and a search for support, marked by hashtags like \#vent, \#edvent, and \#edtt. The use of these tags suggests a community-oriented space for sharing experiences and support.
} \\ \hline
\end{tabular}
\caption{TikTok hashtag profiles. GPT-4-generated summaries of descriptions of videos tagged with moderation evaders. } 
\label{tab:TikTok}
\end{table}

    

\section*{Declarations}

\subsection*{Availability of data and materials}
Due to the sensitive nature of the data, an anonymized version can be made available upon request and after verifying that an IRB approval for the study has been obtained by the requesting institution.

\subsection*{Competing interests}
The authors declare that they have no competing interests.

\subsection*{Funding}
This work was partly funded by the NSF (award \#2331722), by the Swiss NSF (CRSII5\_209250), and by the Defense Advanced Research Projects Agency (DARPA) under Agreement No. HR00112290021.

\subsection*{Authors' contributions}
KL and EF conceptualized the study. LL and CB collected TikTok data, and CB and MDC analyzed all data. KL, CB, and MDC drafted the article. All authors reviewed and approved the final manuscript.

\subsection*{Acknowledgments}

The authors are grateful for the dedicated work of a team of USC students taking on the challenge of data collection and analysis, under the mentorship of Abigail Horn and Joanna Yau. The authors thank Zihao He and Aryan Karnati for collecting Twitter data, and Cinthia Sanchez for providing Reddit data.

\bibliographystyle{IEEEtran}
\bibliography{references_ed,lerman,references}

\end{document}